\documentclass[]{aa}
\usepackage{graphicx}
\usepackage{textcomp}
\usepackage{gensymb}
\usepackage[varg]{txfonts}
\usepackage{color}
\usepackage{appendix}
\usepackage{multirow}
\usepackage{amsmath}
\usepackage[colorlinks,linkcolor=blue,citecolor=blue,linktocpage=true,breaklinks,plainpages=false,urlcolor=blue]{hyperref}

\bibpunct{(}{)}{;}{a}{}{,}
\begin{document}
\title{The influence of NLTE effects in Fe~{\sc i} lines on an inverted atmosphere\\
I. 6301\,\AA{} and 6302\,\AA{} lines formed in 1D NLTE}
\author{H. N. Smitha \inst{1} \and R. Holzreuter \inst{1,2} \and M. van Noort \inst{1} \and S. K. Solanki \inst{1,3}}
\institute{$^{1}$Max-Planck-Institut f\"ur Sonnensystemforschung, Justus-von-Liebig-Weg 3, 37077 G\"ottingen, Germany\\
$^{2}$Institute of Astronomy, ETH Zentrum, 8092 Z\"{u}rich, Switzerland\\
$^{3}$School of Space Research, Kyung Hee University, Yongin, Gyeonggi, 446-701, Republic of Korea\\
\email{smitha@mps.mpg.de}}
\titlerunning{NLTE effects in Fe~{\sc i} lines}
\authorrunning{Smitha et al.}

\abstract
{Ultraviolet (UV) overionisation of iron atoms in the solar atmosphere leads to deviations in their level populations based on Saha-Boltzmann statistics. This causes their line profiles to form in non-local thermodynamic equilibrium (NLTE) conditions. When inverting such profiles to determine atmospheric parameters, the NLTE effects are often neglected and other quantities are tweaked to compensate for deviations from the local thermodynamic equilibrium (LTE). }
{We investigate how the routinely employed LTE inversion of iron lines formed in NLTE underestimates or overestimates atmospheric quantities, such as temperature ($T$), line-of-sight velocity ($v_{\rm LOS}$), magnetic field strength ($B$), and inclination ($\gamma$) while the earlier papers have focused mainly on $T$. Our findings has wide-ranging consequences since many results derived in solar physics are based on inversions of Fe~{\sc i} lines carried out in LTE.}
{We synthesized the Stokes profiles of Fe~{\sc i} 6301.5\,\AA{} and 6302.5\,\AA{} lines in both LTE and NLTE using a snapshot of a 3D magnetohydrodynamic simulation. The profiles were then inverted in LTE. We considered the atmosphere inferred from the inversion of LTE profiles as the fiducial model and compared it to the atmosphere resulting from the inversion of NLTE profiles. The observed differences have been attributed to NLTE effects.}
{Neglecting the NLTE effects introduces errors in the inverted atmosphere. While the errors in $T$ can go up to $13\%$, in $v_{\rm LOS}$ and $B,$  the errors can go as high as $50\%$ or above. We find these errors to be present at all three inversion nodes. Importantly, they survive degradation from the spatial averaging of the profiles.}
%
{We provide an overview of how neglecting NLTE effects influences the values of $T, v_{\rm LOS}$, $B,$ and $\gamma$ that are determined by inverting the Fe~{\sc i} 6300\,\AA{} line pair, as observed, for example, by Hinode/SOT/SP. Errors are found at the sites of granules, intergranular lanes, magnetic elements, and basically in every region susceptible to NLTE effects. For an accurate determination of the atmospheric quantities and their stratification, it is, therefore, important to take the NLTE effects into account.}
 
\maketitle

\begin{figure*}[htbp]
    \centering
    \includegraphics[width=0.7\textwidth]{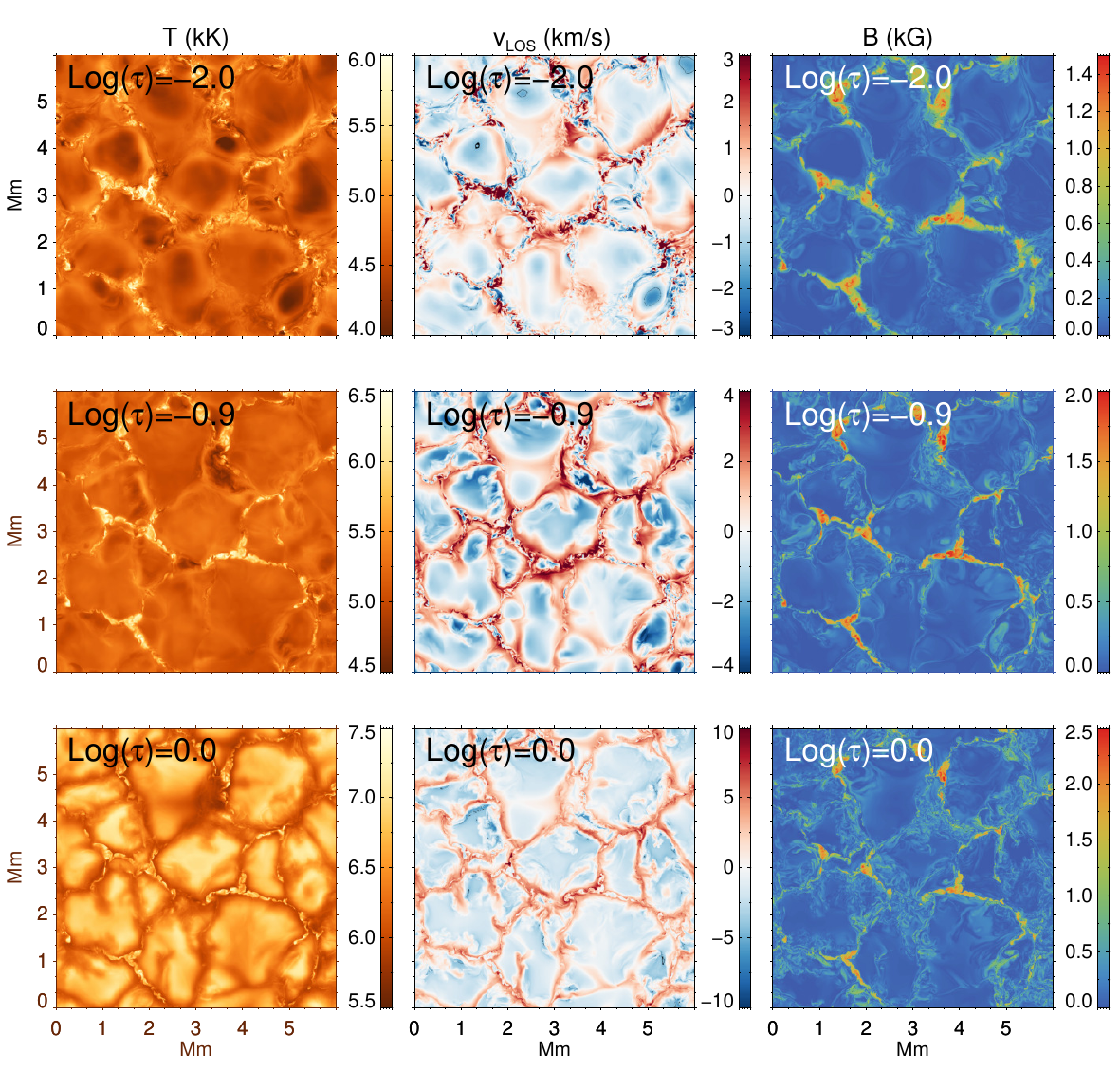}
    \caption{Maps of temperature ($T$), LOS velocity ($v_{\rm LOS}$) and magnetic field strength ($B$) at $log(\tau)=-2.0, -0.9$ and $0.0$ from the 3D MHD cube used for Stokes profile synthesis.}
    \label{fig:maptau}
\end{figure*}

\section{Introduction}
\label{sec:intro}
The photospheric spectral lines of iron at 6301.5\,\AA{} and 6302.5\,\AA{} are among the most widely observed lines for both space and ground-based telescopes. Spectral analysis and the inversion of their Stokes profiles is often carried out based on the assumption of local thermodynamic equilibrium (LTE). The overionization of iron atoms by ultraviolet (UV) radiation, however, results in an underpopulation of the Fe~{\sc i} atomic levels, which leads to deviations from their LTE values. Moreover, with a core formation height between 250\,km -- 300\,km above the photosphere \citep{2010A&A...514A..91G}, their line source function is less than the Planck function, which introduces additional non-LTE (NLTE) effects in the lines themselves. While modeling their spectral profiles, these effects should be taken into account. 

One of the first investigations of the influence of NLTE on iron lines was by \cite{1972ApJ...176..809A}, which was later followed by others, such as \cite{1982A&A...115..104R, 1988ASSL..138..185R, 1988A&A...189..243S, 2001ApJ...550..970S}. Their studies were, however, limited to one-dimensional (1D) radiative transfer. Recently, in a series of papers, \citet[][]{2012A&A...547A..46H, 2013A&A...558A..20H, 2015A&A...582A.101H} investigated the effects of three-dimensional (3D) NLTE radiative transfer on iron lines formed in thin flux tubes and flux sheets in a snapshot of a 3D radiation hydrodynamic (HD) simulation and in a snapshot of a 3D magnetohydrodynamic (MHD) simulation.  In each case, the profiles of the iron lines computed based on the assumption of LTE and 1D NLTE were compared with the 3D NLTE profiles. They reported that in some regions of the atmospheric cube, the NLTE line profiles show line weakening as compared to LTE due to UV irradiation, while a few other regions demonstrated line strengthening due to the departure of the source function from the Planck function, which is in agreement with the findings of \citet[][]{2001ApJ...550..970S}. The NLTE effects were shown to be particularly significant in spatially resolved profiles. 

Accounting for the NLTE effects in iron lines is important  for the determination of solar and stellar atmospheric parameters, including elemental  abundance \citep{2000A&A...359..743A,2005ApJ...618..939S,2012MNRAS.427...50L}. In earlier papers involving the inversion of iron line profiles for the determination of atmospheric models and iron abundances, the NLTE effects have generally been neglected \citep[e.g.,][]{2002A&A...391..331B, 2002A&A...385.1056B, 2008ApJ...678L.157R, 2013A&A...554A..53R, 2013A&A...557A..24V, 2013A&A...557A..25T, 2015A&A...576A..27B}. To our knowledge, there has, so far, been only one exception. Using the NICOLE \citep{2015A&A...577A...7S} NLTE inversion code, \citet[][]{2011A&A...529A..37S} inverted observations from the 6302.5\,\AA{} line taken by the \textit{Hinode} satellite to infer a 3D solar atmospheric model. The NLTE effects were taken into account by using the departure coefficients computed from a 3D HD model. They found that with the NLTE correction, the line cores are better fitted and the inverted atmosphere is less noisy.
 
The current paper aims to investigate the errors introduced in the inferred atmospheres when the NLTE effects in the Fe~{\sc i} 6301.5\,\AA{} and 6302.5\,\AA{} lines are neglected during inversions. For this, we first computed the Stokes profiles from an MHD cube in both LTE and in NLTE. Since we could not at present carry out  a fully self-consistent NLTE inversion of Fe~{\sc i} lines, we decided to test how well LTE inversions (referred to as just an "inversion") of NLTE lines work by comparing them with inversions of LTE lines. That is, we inverted both NLTE and LTE line profiles using an LTE inversion code. To get a rough idea of the errors introduced by neglecting NLTE effects during the inversion, the model atmosphere obtained from the inversion of profiles computed in LTE from the MHD snapshot was used as a reference and the atmosphere recovered from the inversion of NLTE profiles was compared with this reference. Any differences were then classified as errors that the inversion makes by neglecting the NLTE effects.

Earlier studies by \cite{2001ApJ...550..970S, 2013A&A...558A..20H, 2015A&A...582A.101H} focused mainly on the influence of NLTE effects on temperature measurements by analysing the intensity profiles. In the present paper, we also discuss the influence of NLTE effects on the line-of-sight (LOS) velocity and magnetic field diagnostics by comparing the Stokes profiles ($I,Q,U,V$) computed in LTE and NLTE. Here we focus only on the 1D NLTE effects (simply referred to as NLTE) and neglect the effects of horizontal radiative transfer. The influence of horizontal radiative transfer will be investigated in a follow-up paper.

Finally, we tested whether the spatial averaging of the Stokes profiles influences the accuracy of the atmosphere derived in LTE. For this, we rebinned the profiles to match the spatial scales resolved by \textit{Hinode} and inverted the profiles as before. We find that because the NLTE effects indeed appear to survive such spatial degradation, it is important to account for them to infer atmospheric quantities accurately. 

\begin{figure*}[htbp]
\centering
\includegraphics[width=0.80\textwidth]{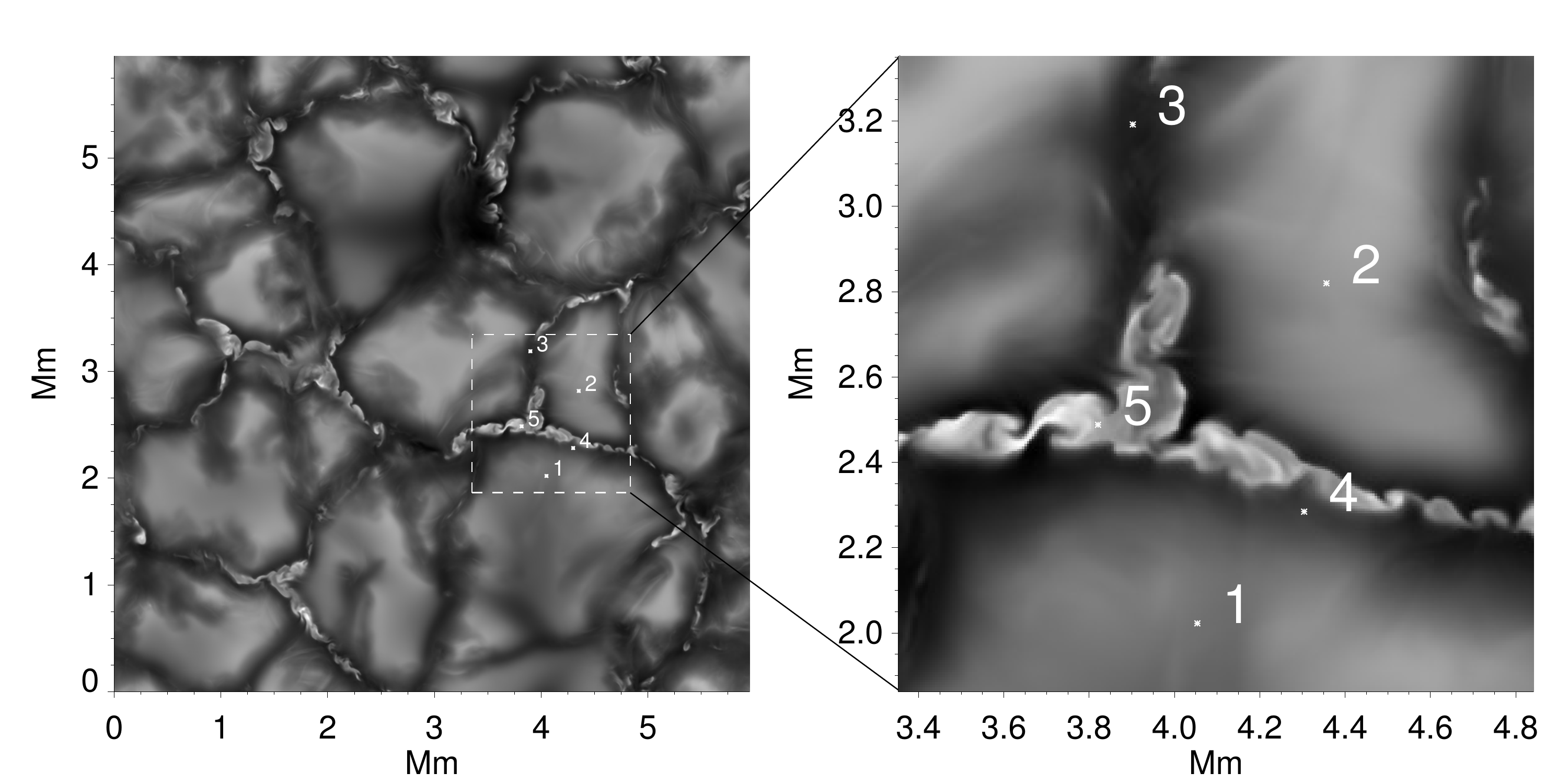}
\caption{Continuum image of the analyzed snapshot. Dashed white box represents the area chosen for 3D NLTE computation of the iron lines, the results of which will be discussed in a future work. A blow-up of the region covered within the white box and the five points are shown on the right. For detailed analysis of the Stokes profiles, we picked five spatial positions numbered 1-5 within this box.}
\label{fig:cont1}
\end{figure*}

\section{Model atmosphere and Stokes profiles synthesis}
\label{sec:atmos}
A Snapshot of a realistic 3D MHD model atmosphere generated by the MURaM code \citep{2005A&A...429..335V} was used for the synthesis of the Stokes profiles. The cube extends 6 Mm $\times$ 6 Mm $\times$ 2 Mm in the x, y, and z directions, respectively, with a resolution of 5.82 km $\times$ 5.82 km $\times$ 7.85 km. The cube represents the quiet Sun with mixed magnetic polarities. 

Maps of the temperature ($T$), LOS velocity ($v_{\rm LOS}$) and magnetic field strength ($B$) at $log(\tau)=-2.0, -0.9,$ and $0.0$ from the 3D MHD cube are shown in Figure~\ref{fig:maptau}.  In the rest of the paper, by $log(\tau)$ we mean $log(\tau_{5000})$. We choose these three $\tau$ surfaces because the nodes for inversions are placed there, as will be discussed in Section~\ref{sec:inversion}. As we move higher up in the atmosphere, from $log(\tau)=0$ to $-2.0$, the temperature maps clearly change from normal to reverse granulation, with some granules still hotter than others at $log(\tau)=-2.0$, but clearly cooler than the intergranular lanes. 
As the height of the $log(\tau)=-0.9$ surface is approximately midway through this transition, the distribution of the temperature across the granules appears more homogeneous. Regions with a high magnetic field strength stand out as hot regions between the granules, with temperatures much higher than the average in intergranular lanes. 
A narrowing of the intergranular lanes and increase in the convective velocities (upflows and downflows) with depth can be observed.

The Stokes profiles of the 6301.5\,\AA{} and 6302.5\,\AA{} iron lines were synthesized in 1D using the RH code \citep{2001ApJ...557..389U} in both LTE and NLTE, at $\mu=1$. By 1D computations, we really mean 1.5D, that is 1D radiative transfer carried out along each column of the MHD cube, while neglecting all 3D radiative transfer effects in the profile synthesis. All calculations were done in the same way as given in \citet[][]{2015A&A...582A.101H}. Similar to the computations in that paper, we neglect the influence of NLTE effects on the concentration of H- ions, which can affect the continuum \citep{2010A&A...517A..48S}. The full Stokes profiles were computed using the field-free approximation \citep{1969SoPh...10..268R}. The logarithmic iron abundance was taken to be 7.50 on a scale on which hydrogen has an abundance of 12. 

We used the same iron atomic model that was used in \citet[][]{2013A&A...558A..20H, 2015A&A...582A.101H}. The model contains 23 atomic levels coupled by 33 bound-bound transitions and 22 bound-free transitions. The bound-free transitions represent the coupling of atomic levels in each term in Fe~{\sc i} to its parent term in Fe~{\sc ii} through photoionization. For the photoionization cross-sections, hydrogen-like approximations have been used. The missing iron line opacity in the UV was approximated
by enhancing the opacities in the relevant wavelength range, which is known as opacity fudging, as proposed by \cite{1992A&A...265..237B}. \citet[][their Section~3.2]{2012A&A...547A..46H} have tested the effect of such an opacity fudging on the synthesized Fe~{\sc i} lines. They find that the 6301.5\,\AA{} line is less sensitive to the opacity fudging compared to the 5250\,\AA{} line. Removing the opacity fudging will likely enhance the NLTE effects in these lines possibly to unrealistically large values.

The continuum intensity map at 6300\,\AA{} is shown in Figure~\ref{fig:cont1}. To facilitate the comparison of the different cases, we choose five representative spatial points within the white box, inside a granule (pos. 1 and 2), in an intergranular lane (pos. 3), intergranular lane next to a magnetic element (pos. 4) and in the middle of a strong magnetic element (pos. 5), the last two of which, following the convention in \citet[][]{2015A&A...582A.101H}, may be referred to as a flux sheet and a flux tube, respectively. Although we discuss the full maps of the different atmospheric quantities obtained from inversions, at these five spatial locations we also  discuss the Stokes profiles in detail.

\section{Inversion of Stokes profiles}
\label{sec:inversion}
The Stokes profiles computed assuming LTE and NLTE conditions were inverted using the 1D LTE inversion code  Stokes-Profiles-INversion-O-Routines, \citep[SPINOR,][]{2000A&A...358.1109F}, which makes use of the STOPRO routines \citep{1987PhDT.......251S}. The code assumes a strongly simplified model of the atmospheric stratification, consisting of a spline representation of the temperature, magnetic field strength, inclination ($\gamma$), azimuth, and LOS velocity as a function of optical depth. The splines are controlled by the values of each atmospheric quantity at three points, referred to as nodes, placed at strategically chosen optical depth values throughout the formation region of the spectral lines to be inverted. To avoid any arbitrariness as much as possible, for the inversions presented in this paper, the nodes were placed at $log(\tau)=-2.0, -0.9$ and $0.0$, close to the optimum found by \citet[][]{2016A&A...593A..93D}. No microturbulent velocity was used in fitting the Stokes profiles since none was used to synthesize them in the first place. The SPINOR code only returns the values of the atmospheric quantities at the positions of the three nodes and the comparison between the atmospheres from the inversion of LTE and NLTE profiles are, therefore, shown only on these three iso-$(\tau)$ surfaces.

\begin{figure*}[htbp]
    \centering
    \includegraphics[width=0.49\textwidth]{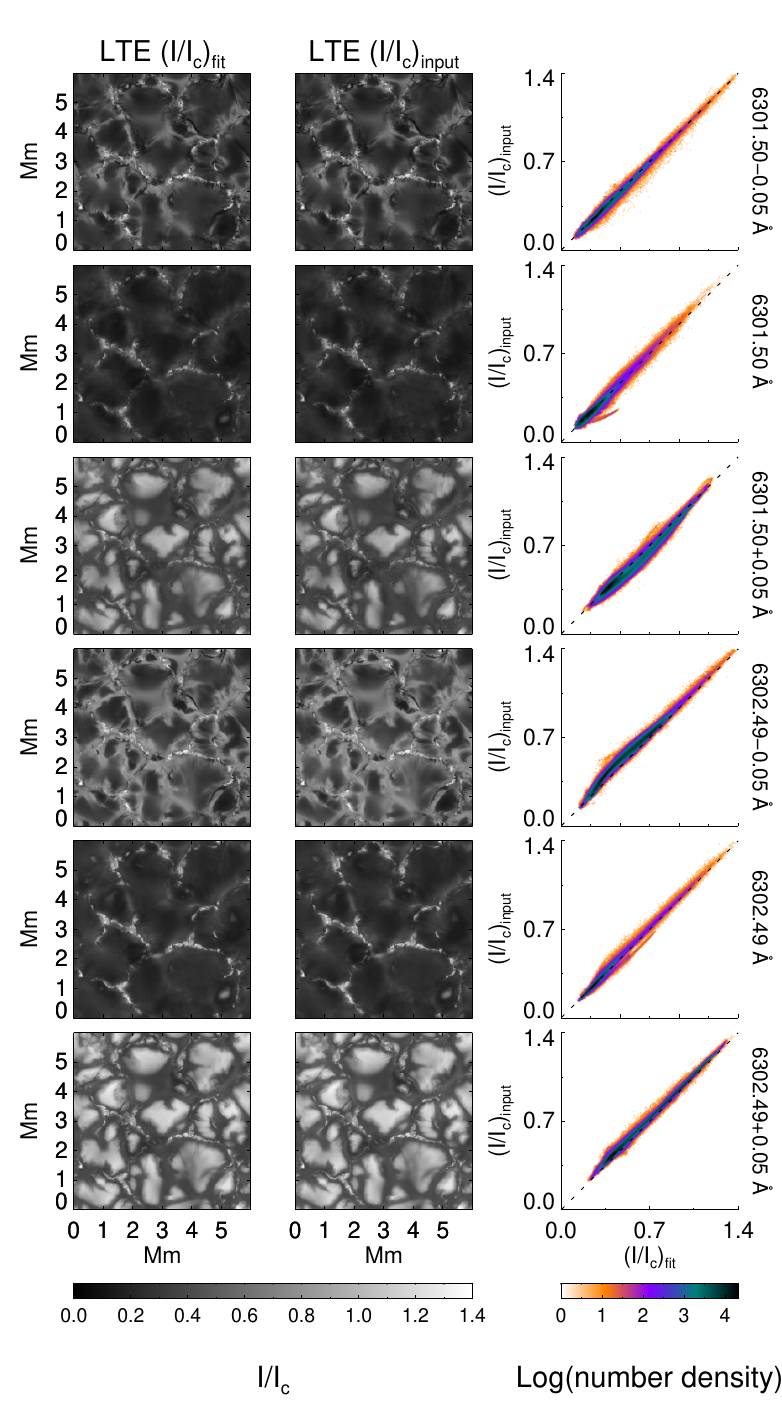}
    \includegraphics[width=0.49\textwidth]{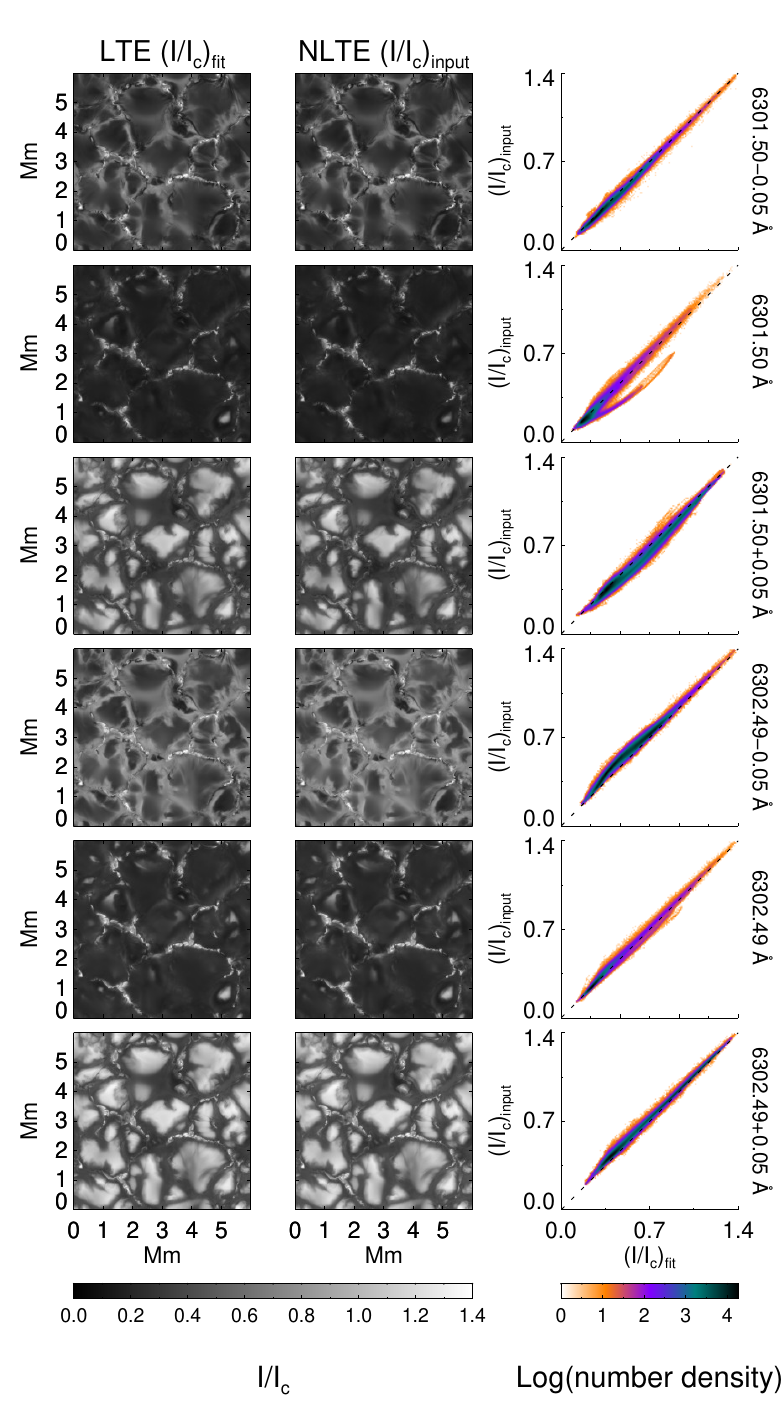}
        \caption{Intensity maps from the LTE profiles generated by the inversion code (\textit{first column}) to fit the input LTE maps in the \textit{second column}, at six different wavelength points. They are at the centers of the two lines and at $\pm 0.05$\,\AA{} from their line centers. The two maps are compared using scatter density plots in the \textit{third column}. The other three columns on the right indicate the same but for NLTE input profiles.}
    \label{fig:icomp}
\end{figure*}

In Figure~\ref{fig:icomp}, we show the input intensity maps and the best fits at six wavelength points; specifically, at the centers of the two lines, and $\pm 0.05$\,\AA{} on either side of the line centers. The plotted intensities are normalized to $I_c$, the continuum intensity, spatially averaged over the surface of the whole cube. The accuracy of the fit is demonstrated using the scatter density plots. The intensity maps look almost identical in both cases at all six points. The scatter plots are narrow, with the majority of the data points lying along the diagonal. The same is true for the Stokes $Q, U$ and $V$ profiles (figures not shown). The low scatter of the values around the diagonal shows that the inversion code is able to fit the LTE and NLTE profiles equally well and, therefore, any difference in the two (LTE and NLTE) input profiles is translated by the inversion into differences in the atmospheric parameters. 

\begin{figure*}[htbp]
\begin{center}
\begin{minipage}{0.46\textwidth}
\centering
\includegraphics[width=0.85\textwidth]{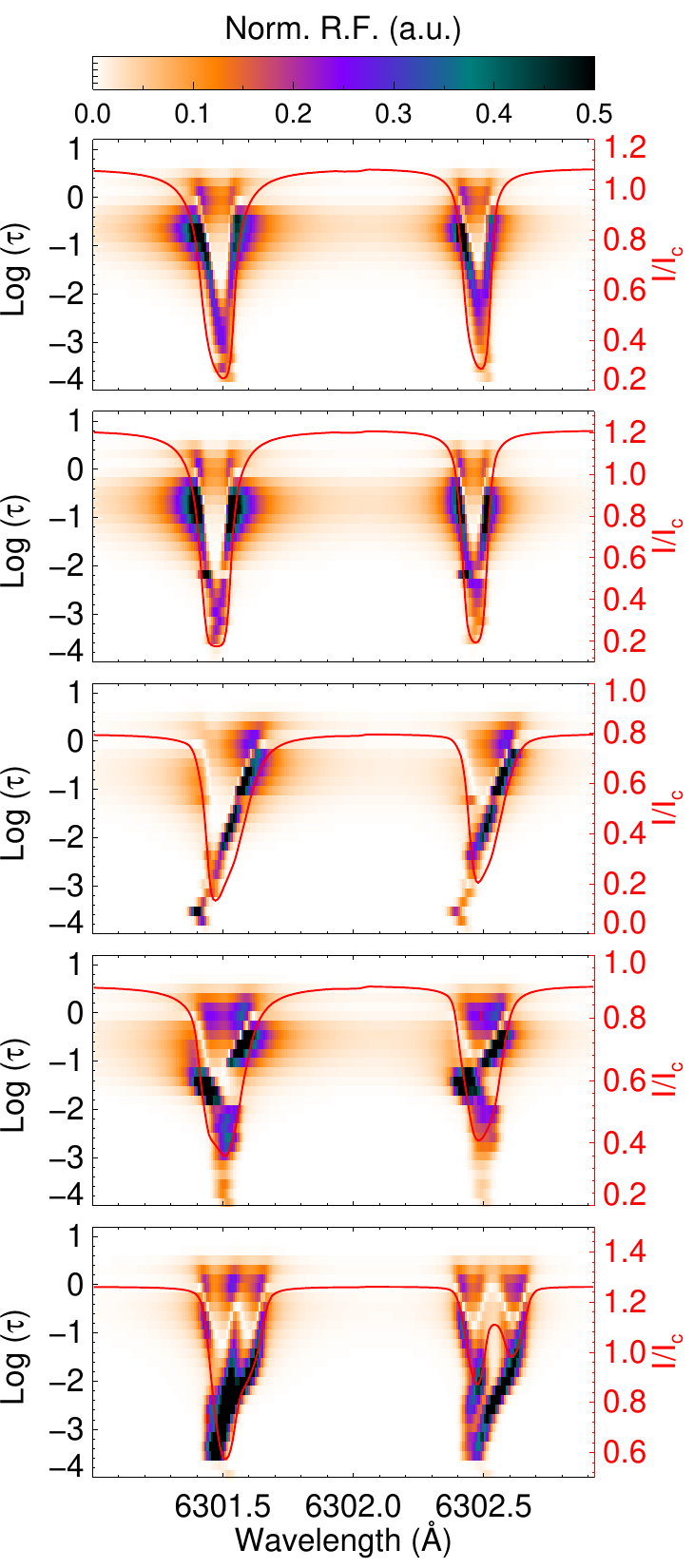}
\end{minipage}
\begin{minipage}{0.50\textwidth}
\centering
\includegraphics[width=1.01\textwidth]{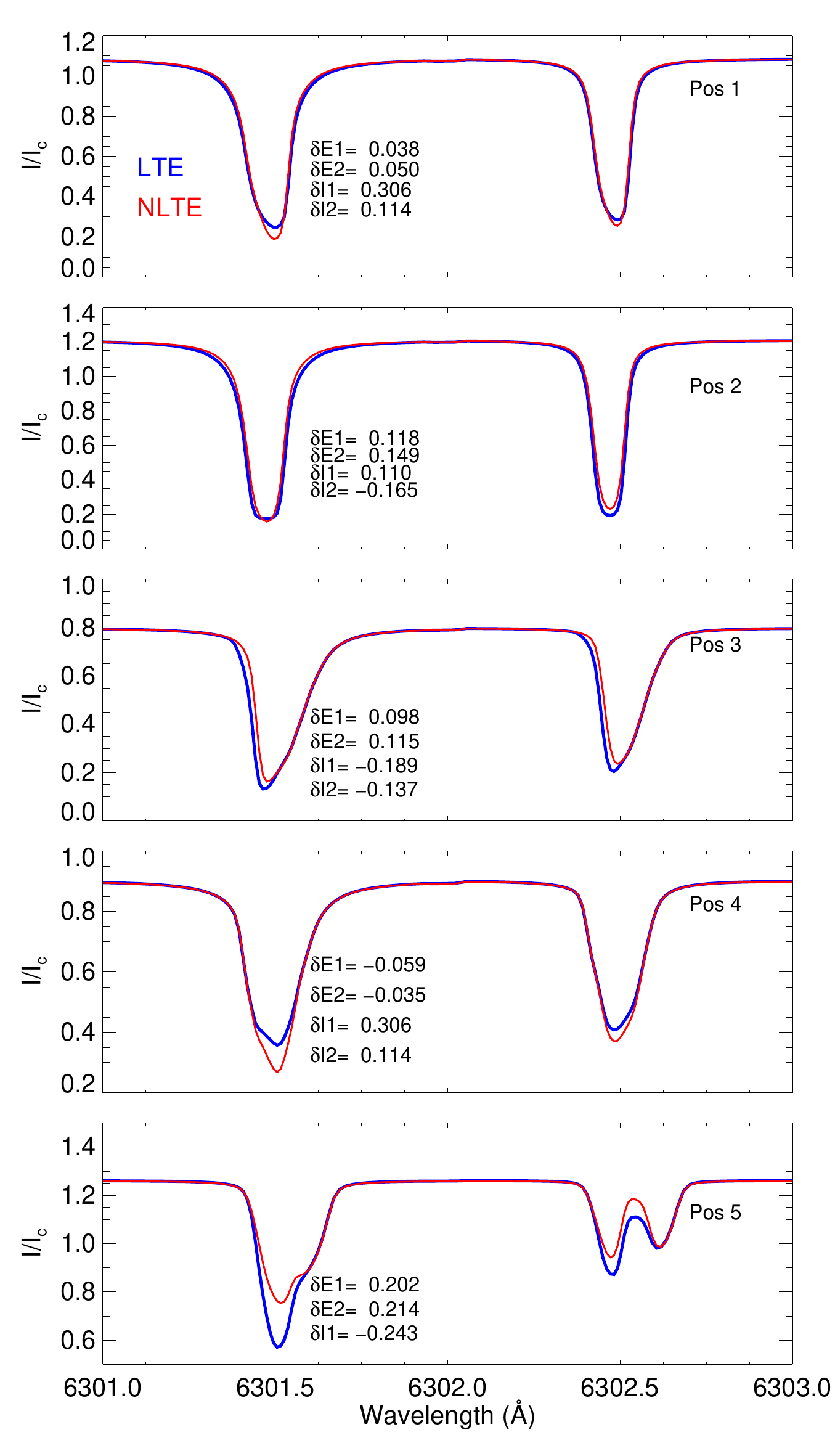}
\end{minipage}
\caption{\textit{Left:} Response functions of intensity to temperature at the five chosen spatial positions (see Figure~\ref{fig:cont1}). The response functions are normalized to unity but are displayed here in the range 0--0.5 for improved contrast. The LTE intensity profiles in red color are overplotted on these 2D maps . \textit{Right:} Comparison between LTE and NLTE intensities at the five spatial positions marked in Figure~\ref{fig:cont1}. $\delta I$ and $\delta E$ are the relative differences in intensity and equivalent width defined in Equations~\ref{eqn:int_ew}. The subscripts 1 and 2 refer to the 6301\,\AA{} and 6302\,\AA{} lines, respectively. At pos 5, $\delta I_2$ is not indicated as the line profiles are Zeeman split.}
\label{fig:rf_int}
\end{center}
\end{figure*}

\begin{figure*}[htbp]
    \centering
    \includegraphics[width=0.38\textwidth]{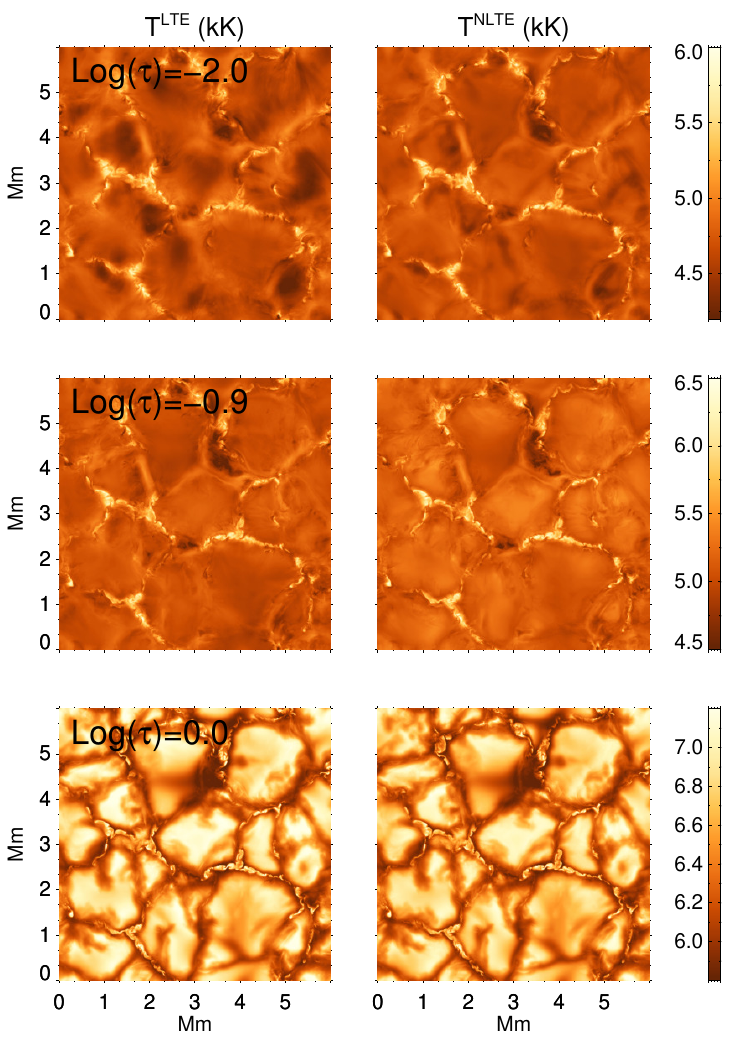}
    \includegraphics[width=0.595\textwidth]{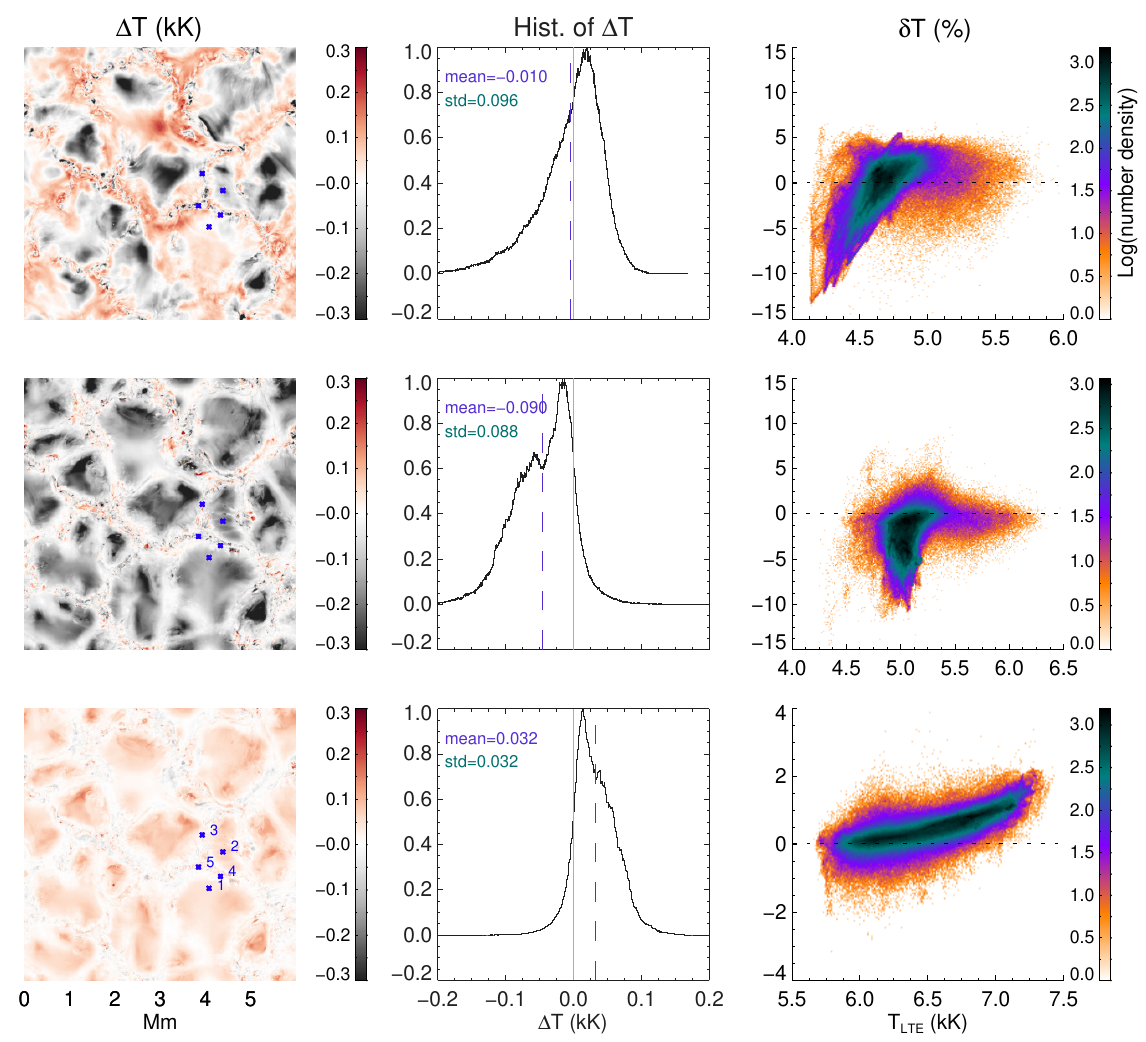}
    \caption{Comparison between the temperature maps from the inversion of LTE (\textit{first column}) and NLTE (\textit{second column}) Stokes profiles at the three nodes $log(\tau)= -2.0$ (\textit{top row}), $-0.9$ (\textit{middle row}), and $0.0$ (\textit{bottom row}). Difference in temperature between the first two columns ($\Delta T$) and normalized histograms of these differences are shown in the \textit{third} and \textit{fourth} columns. The mean and standard deviation of the histogram distributions are indicated (in units of kK). The position of mean is indicated by the vertical dashed line plotted over the histogram. In the \textit{fifth} column, we show the relative difference $\delta T$ computed using Equation~\ref{eqn:rel_diff}, expressed in percentage,  as a function of temperature from the inversion of LTE profiles which is used as reference.}
    \label{fig:temp}
\end{figure*}

\section{Comparison of the inverted atmospheres}
\label{sec:atm_comp}

An ideal way to quantify the quality of the model atmospheres returned by the inversions would be to compare them with the MHD cube. For Milne-Eddington inversions, one could follow a strategy similar to that of \citet[][]{2014a&a...572a..54b} and compare the two. However, when we have a stratified atmosphere, comparison with the MHD cube is complicated and there is no unique way to do so. The method used for such a comparison would introduce additional uncertainties. To avoid this, we first inverted the LTE profiles in LTE. This case is consistent so that the results should be reasonably correct (although grossly simplified compared with the original MHD snapshot). We used this atmosphere as the fixed reference and called it the "reference model".  We then inverted the NLTE profiles in LTE and referred to the atmosphere obtained as the "test model". This latter case is not self-consistent and differences between the reference model and test model are likely due to the NLTE effects. A similar approach was used in \cite{2012A&A...548A...5V} but the input was re-synthesized from the fitted atmosphere, which is not done in the present paper

The reference and test models are compared in two ways, one simply using the difference
\begin{equation}
    \Delta x = x^{\rm LTE}-x^{\rm NLTE},
    \label{eqn:abs_diff}
    \end{equation}
and the other using the relative difference defined as
    \begin{equation}
    \delta x = \frac{x^{\rm LTE}-x^{\rm NLTE}}{x^{\rm LTE}},
    \label{eqn:rel_diff}
    \end{equation}
where $x$ represents an atmospheric quantity, such as temperature, velocity, magnetic field strength, or inclination.

The two spectral lines at 6301.5\,\AA{} and 6302.5\,\AA{} are formed approximately 100 km apart in the middle to upper photosphere \citep{2010A&A...514A..91G}, with the 6301.5\,\AA{} line forming somewhat higher than the 6302.5\,\AA{} line, but with a smaller Land\'{e} g-factor. Due to its formation height, the 6301.5\,\AA{} line is more affected by the NLTE conditions than the 6302.5\,\AA{} line. Since the two lines are inverted together, the NLTE sensitivity of both lines contributes towards shaping the final inverted atmosphere, making the interpretation more complex, especially with a stratified atmosphere.

Different parts of a spectral line, from continuum to the line core, are formed at different heights in the atmosphere. So the contributions to the inferred atmosphere at the three nodes come from different parts of the spectral line. This information is best captured by the so-called response functions, which tells us how the values of the Stokes parameters at each wavelength point responds to a change of a physical quantity in the atmosphere at a particular depth. In Figure~\ref{fig:rf_int}, we show the LTE response functions of intensity to temperature at the five chosen spatial points. The value of the response functions, normalized to unity and represented by a color value, is plotted as a function of the spectral range and the optical depth $log(\tau)$. Over-plotted on this are the LTE intensity profiles at the five locations. Each spectral point has a non-zero response over a range of optical depths. For example, the response function at the line core is strongest at lower optical depths (larger heights) in the atmosphere and decreases with depth. As a first approximation, we assume that the contribution to the top node $log(\tau)=-2.0$ comes mainly from the line core. Such an assumption will help us explain the differences in temperature maps by correlating them to the differences in the line core intensities of the LTE and NLTE profiles. In the right panels of the same figure, we also compare the LTE and NLTE intensity profiles at the five points. We compute the relative differences in intensity (or residual intensity) and equivalent widths similar to Equations 1 and 2 of \citet[][]{2015A&A...582A.101H}. The expressions are repeated here for convenience,
\begin{eqnarray}
    \delta I = \frac{I^{\rm LTE}-I^{\rm NLTE}}{I^{\rm NLTE}};
     \delta E = \frac{EW^{\rm LTE}-EW^{\rm NLTE}}{EW^{\rm NLTE}},
    \label{eqn:int_ew}
\end{eqnarray}
where $I^{\rm LTE, NLTE}$ is the minimum intensity of an LTE/NLTE profile at a specific spatial location and $EW^{\rm LTE, NLTE}$ is the equivalent width. A $\delta I < 0$ or a $\delta E >0$ corresponds to NLTE weakening of the line. An overview of the variation in $\delta E$ for the 6301.5\,\AA{} line across different atmospheric structures from an MHD cube, like granules, intergranular lanes, magnetic elements (flux tubes and flux sheets) etc., can be found in Figure~3 of \citet[][]{2015A&A...582A.101H}. In Figure~\ref{fig:rf_int}, we indicate the values of $\delta I$ and $\delta E$ only at the representative five spatial points computed from the intensity profiles, for both the 6301.5\,\AA{} and the 6302.5\,\AA{} lines. 

\subsection{Temperature}
\label{sec:temp}
\subsubsection{Top and central nodes}
\label{sec:temp_nodes12}

The temperature maps from the reference and test models at $log(\tau)=-2.0, -0.9,$ and $0.0$ are compared in Figure~\ref{fig:temp} (\textit{first and second columns}). We first discuss the maps in the top two nodes ($log(\tau)=-2.0$ and $-0.9$) and consider the bottom node ($log(\tau)=0.0$) separately in Section~\ref{sec:temp_node0}. In Figure~\ref{fig:temp}, the temperature maps from the test model look quite similar to those of the reference model. However, the difference map in the \textit{third column} indicates that the $\Delta T$ is mostly negative (i.e., the temperature in the test model is higher than in the reference model) in the granules and can reach values as high as 300\,K in some patches. The higher temperature in the test model is also reflected in the histogram of $\Delta T$ (\textit{fourth column}), which displays an increased skewness towards negative values. The mean and standard deviation of the histogram distribution (first and second numbers on the left side, respectively) are less than 100 K. The density plots of the relative difference $\delta T$, computed using Equation~\ref{eqn:rel_diff}, are shown in the last column. In the top node, we find $\delta T$ reaching up to -13\% in regions where $T^{\rm LTE}$ $<$ 5000\,K, which look like reverse granulation.
In regions of higher temperature ($T^{\rm LTE} > 5000 K$), $\delta T$ can be positive or negative depending on the temperature gradients in those regions, and it remains below 5\%. Overall, the density of points is within $\pm 5\%$ in the top node.  In the middle node, $\delta T$ is again prominently negative in the granules reaching up to 10\% while in the intergranular lanes it is less than 5\%. 

\begin{figure*}[htbp]
\centering  
\includegraphics[width=0.69\textwidth]{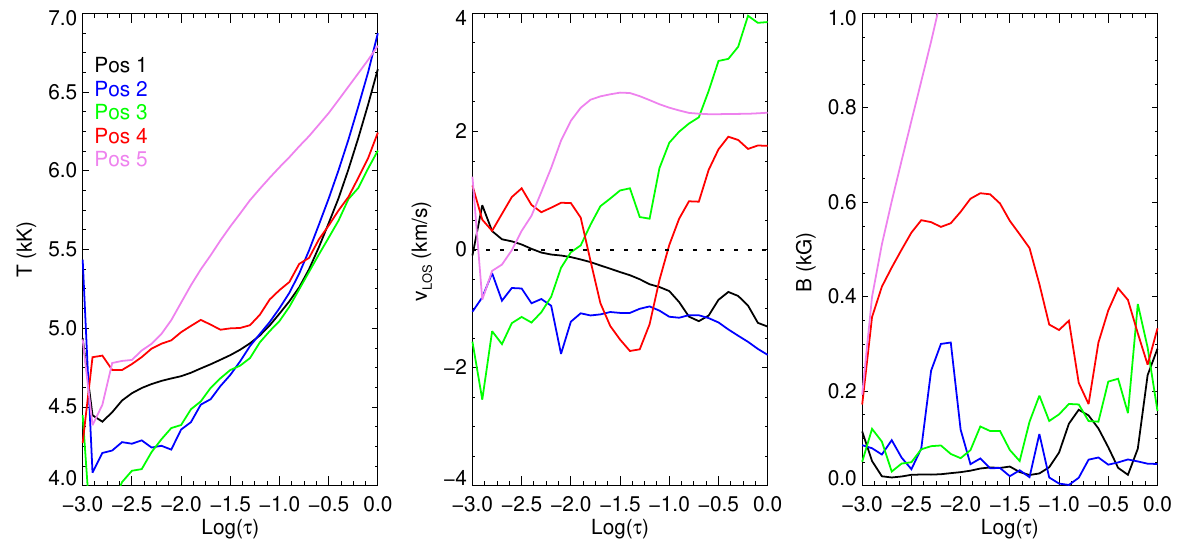}
\caption{Variation in temperature (\textit{first panel}), LOS velocity (\textit{second panel}), and magnetic field strength (\textit{third panel}) with $log(\tau)$ in the MHD cube at the five spatial positions marked in Figure~\ref{fig:cont1}.}
\label{fig:maptau_grad}
\end{figure*}

\begin{table*}[htbp]
    \centering
        \caption{Errors in the  measurement of atmospheric quantities at the five spatial points when NLTE effects are neglected.} 
    \begin{tabular}{ccccccc}
    \hline
     & $log(\tau)$ & Pos 1 & Pos 2 & Pos 3 & Pos 4 & Pos 5  \\
    \hline
    \multirow{3}{*}{$\Delta T$}\\
        &-2.0 & 16.8 (4721) & -158.5 (4490) & -126.2 (4404) & 121.8 (4919)& -130.7 (5299) \\
    (K) &-0.9 & -108.5 (5036) & -217.0 (5087) & -78.3 (5005)& 8.6 (5097) & -45.9 (5963) \\
        &  0.0 & 35.9 (6718) & 80.2 (6968) & 36.9 (6096)& 0.5 (6295)& 29.9 (6761) \\
         \hline
    \multirow{3}{*}{$\Delta v_{\rm LOS}$}\\
        &-2.0 & 0.08 (-0.27) & -0.01 (-1.1) & -0.53 (-0.1) & 0.55 (0.18) & -0.55 (1.6) \\
(km/s)  &-0.9 & 0.07 (-0.64) & 0.03 (-1.1)& -0.35 (1.4) & -0.59 (-0.7) & 0.02 (2.8) \\
        & 0.0 & -0.15 (-1.75)& 0.10 (-1.9)& -0.09 (4.1) & 0.66 (3.7)& 0.15 (1.5)  \\
         \hline
             \multirow{3}{*}{$\Delta B$}\\
        &-2.0 & 0.0 (10) & 112.5 (122) & -57.14 (67) & -170.1 (577)& -327.8 (1190) \\
    (G) &-0.9 & 25.7 (131) & -14.1 (10) & 26.4 (180) & 454.0 (571)& 1.5 (1856) \\
        & 0.0 &-84.7 (255)& -23.7 (224) & -62.4 (234)& -203.8 (217)& -123.2 (1720) \\
         \hline
                      \multirow{3}{*}{$\Delta \gamma$}\\
        &-2.0 & -33.6 (70.2) & -30.2 (95.4) & -38.9 (63.1) & 22.3 (179.5)& -1.5 (162.9) \\
    (deg) &-0.9 & -2.7 (103.1) & -6.7 (99.8) & -5.0 (126.5) & 10.0 (128.9)& -4.0 (175.6) \\
        & 0.0 & 3.0 (107.7)& 4.5 (94.0) & 5.1 (71.2)& 1.6 (1.6)& -1.6 (158.2) \\
         \hline
    \end{tabular}
    \tablefoot{The table shows errors in the  measurement of temperature, LOS velocity, magnetic field strength, and inclination at the five spatial points when NLTE effects are neglected. The numbers indicate simple difference defined in Equation~\ref{eqn:abs_diff}. A positive difference indicates that the inversion of NLTE profiles returns a value smaller than in the reference atmosphere which is from the inversion of LTE profiles. The numbers in parentheses are the values from the reference atmosphere}
\label{tab:table1}
\end{table*}

Now we consider the temperature differences at the 5 spatial points in detail. The differences in temperature between the reference and test models at the five points in the three nodes are given in Table~\ref{tab:table1}. The temperatures in the reference model are indicated in parentheses. From Figure~\ref{fig:rf_int}, the $\delta E$ is positive in the middle of the granules (pos. 1 and 2), in intergranular lanes (pos. 3) and in the middle of a magnetic concentration (pos. 5). It is negative only in the intergranular lane next to a magnetic element (pos. 4), as discussed also in \citet[][]{2015A&A...582A.101H}. The weakening of the NLTE line relative to LTE ($\delta E$ > 0) is due to the opacity deficit resulting from the UV overionization of iron atoms. When this weak NLTE line is fit by the inversion code using an LTE line, the temperature inferred is higher than that inferred from the inversion of an LTE line profile. However, in some cases, the NLTE line can also be stronger than its LTE version. This happens due to the departure of source function from the Planck function. When this effect dominates over the line weakening caused from the UV overionization \citep[for a detailed discussion, see][]{2001ApJ...550..970S}, we observe line strengthening due to NLTE effects. For example, in the top node, $\Delta T$ > 0 in pos. 1 but negative in pos. 2 (see \textit{third column} in Figure~\ref{fig:temp} and also in Table~\ref{tab:table1}). In other words, although both pos. 1 and pos. 2 are in the granules, temperature from the inversion of NLTE lines is higher than from the inversion of LTE lines only in pos. 2 and exactly the opposite is observed at pos. 1. 
To explain this further, we plot the temperature profiles at these locations from the MHD cube as a function of $log(\tau)$ in Figure~\ref{fig:maptau_grad}. The temperature stratification is quite similar at both pos.1 and pos.2 up to $log(\tau)=-1.0$. In the higher layers, the temperature gradient is smaller at pos.1 compared to pos.2. The smaller gradient leads to the strengthening of the line formed in NLTE compared to the LTE line. The residual intensity, $\delta I$,  for both lines at pos.1, indicated in Figure 4, are positive and large, 30\% for 6301.5\,\AA{} line and 11\% for 6302.5\,\AA{} line. This indicates line strengthening. When the inversion code fits both these lines in LTE, the temperature from fitting the NLTE line will be lower than that from fitting the LTE line, resulting in a positive $\Delta T$ at pos.1. Such a line strengthening at several locations in the model atmosphere was observed also by \citet{2012A&A...547A..46H,2013A&A...558A..20H,2015A&A...582A.101H}.

 In the intergranular lanes, \citet[][]{2013A&A...558A..20H} predicted that applying LTE inversions would underestimate the temperature by 100 K - 200 K \citep[also discussed by][]{2001ApJ...550..970S}. This conjecture is in agreement with the difference image at the top node (\textit{third column}, Figure~\ref{fig:temp}). For example, at pos. 3, the NLTE intensity profiles are shallower than in LTE, $\delta I<0$ for both the lines (Figure~\ref{fig:rf_int}), and $\Delta T$ is negative (Table~\ref{tab:table1}). In intergranular lanes next to magnetic elements, for example pos. 4, the temperature gradients are small, NLTE effects are not as strong, and $\delta E$ < 0, $\delta I$ >0. \citet[][]{2013A&A...558A..20H} discuss that in the border of granules, the LTE temperature can be 200 K-400 K higher than in NLTE. This effect is seen in the top node, where along the boundaries of the granules, the difference image shows the temperature in reference model to be higher than in the test model (see Figure~\ref{fig:temp}, \textit{third column}). Since in both \citet[][]{2013A&A...558A..20H} and  \cite{2001ApJ...550..970S}, these predictions were mainly based on the differences in the LTE and NLTE line core depths, they show a good correlation with our results in the top most node.

From Figure~\ref{fig:temp}, we see that in the middle node, the temperature in the test model is higher in all the granules compared to the reference model.  The histogram of $\Delta T$ has a mean value around 90\,K with a small secondary peak on the negative side which is due to the negative $\Delta T$ from the granules. From Figure~\ref{fig:rf_int}, the response at $log(\tau)=-0.9$ is strongest near the line wings.  Differences between the NLTE and the LTE profiles in the line wings result in differences in the equivalent widths which manifest themselves as differences in temperature. In the intergranular lanes, the $\Delta T$ is much smaller as compared to that in the granules. In the intergranular lanes next to magnetic elements (e.g., pos. 4) or in the boundaries of the granules the difference is close to zero. 

\begin{figure}[htbp]
\centering 
\includegraphics[width=0.49\textwidth]{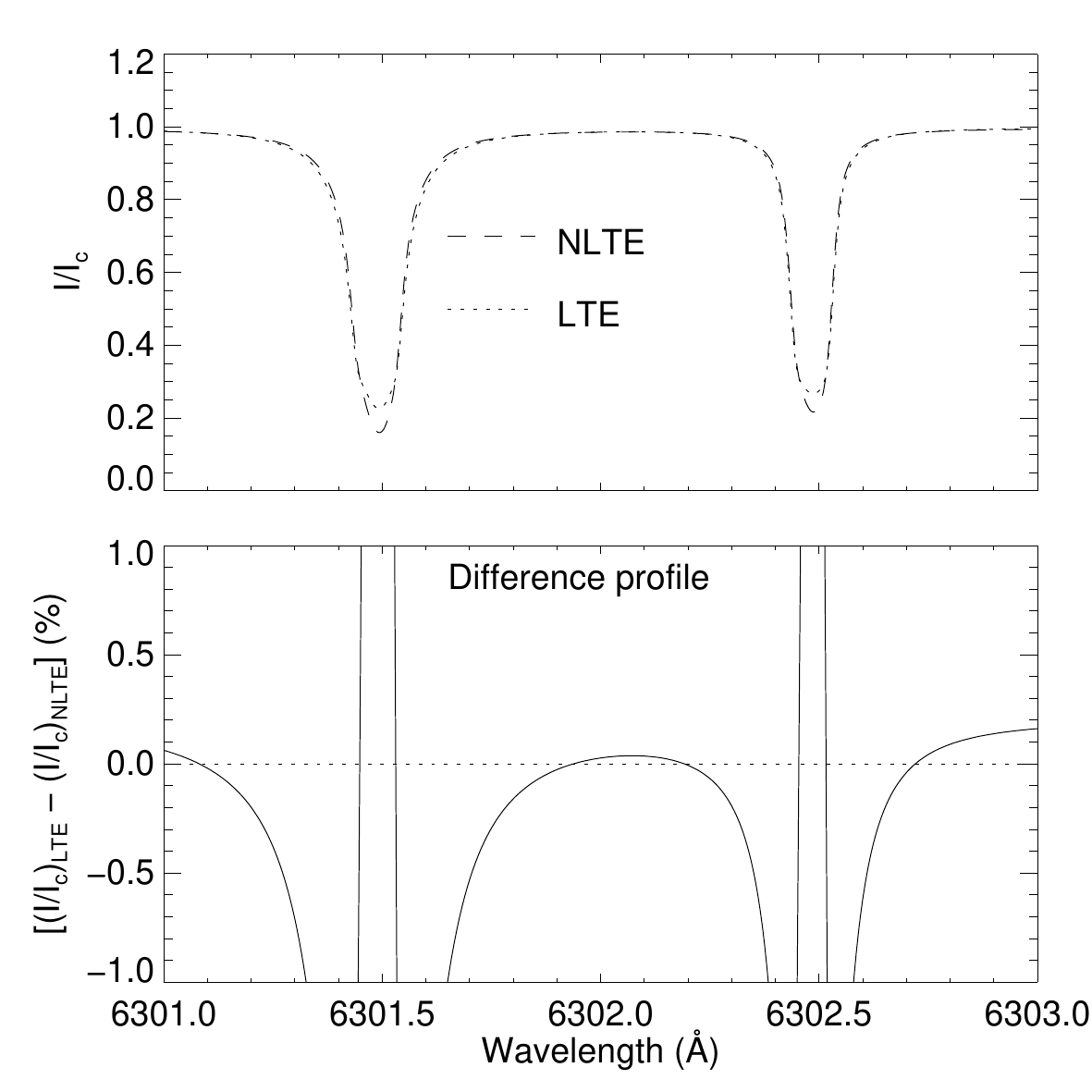}
\caption{\textit{Top panel:}  input LTE and NLTE intensity profiles at a pixel in a granule. \textit{Bottom panel:} difference of the two intensity profiles.}
\label{fig:cont_diff}
\end{figure}

\subsubsection{Bottom node}
\label{sec:temp_node0}
At $log(\tau)=0.0$, the temperature in the reference model is slightly higher than in the test model (Figure~\ref{fig:temp}, \textit{bottom} row). In principle, the inversion code  should return the same temperature in both cases since the continuum intensity is computed strictly in LTE, as in \citet[][]{2015A&A...582A.101H}. However, in between and away from the two spectral lines, the difference between the NLTE and LTE intensity profiles is non-zero and positive. An example from a pixel in the granule is shown in Figure~\ref{fig:cont_diff}. This indicates that within the spectral window used for synthesis, the profiles do not yet reach the real continuum.  The inversion code is thus constraining the temperature at $log(\tau)=0$ by fitting the intensity in the far wings of the spectral lines and not from the continuum intensity. From Figure~\ref{fig:temp}, this results in a difference of about $50-100$\,K in the granules with a relative difference $\delta T$ of less than 2\% . 
\begin{figure*}[htbp]
\centering
 \includegraphics[width=0.38\textwidth]{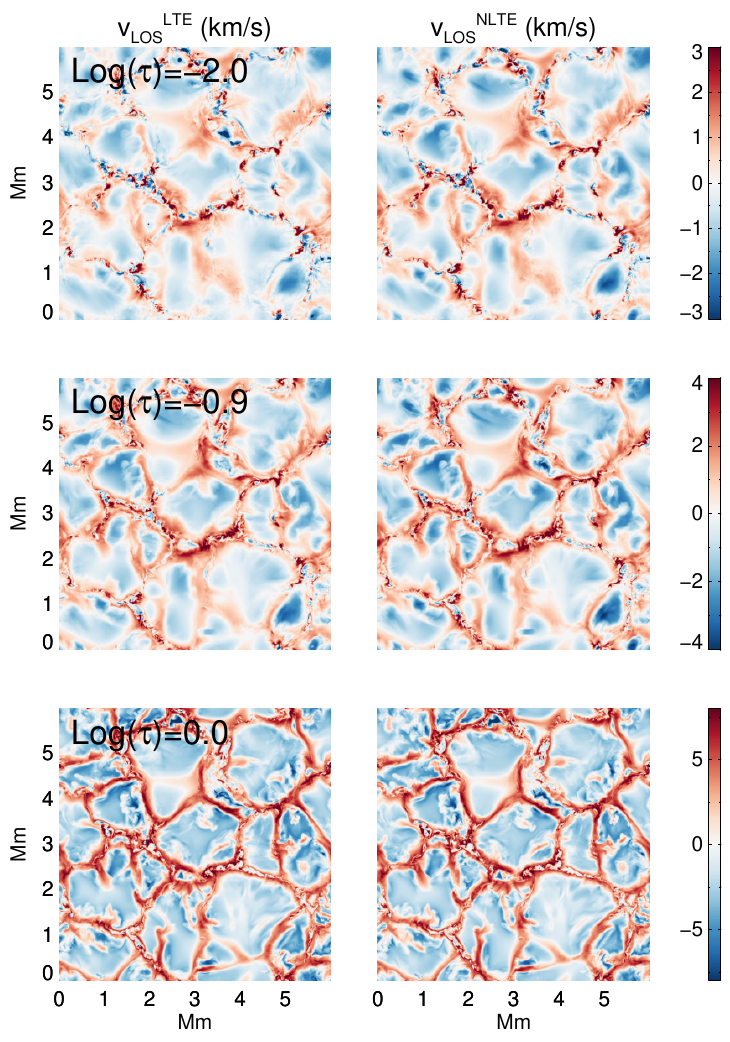}
    \includegraphics[width=0.595\textwidth]{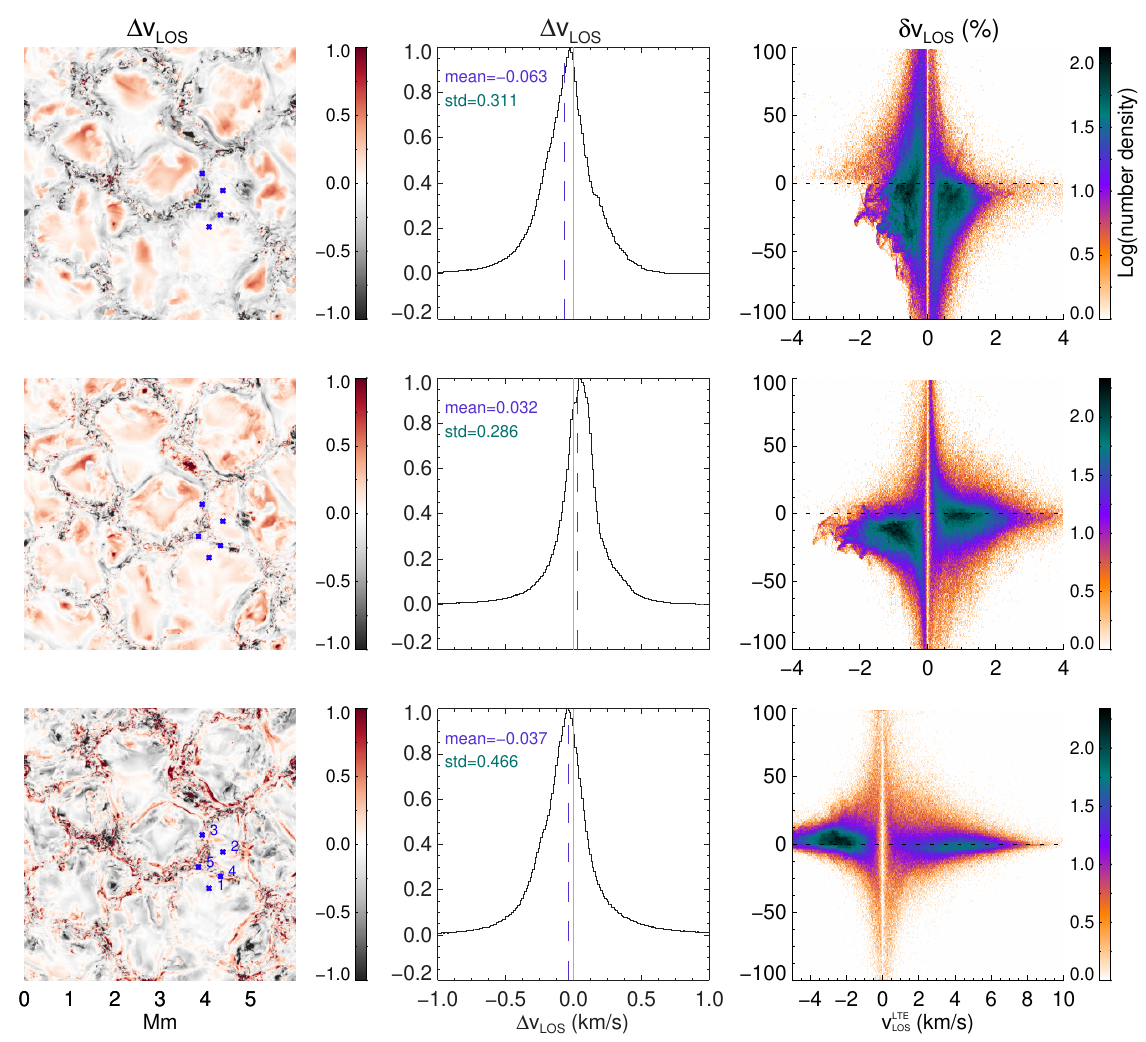}
\caption{Same as Figure~\ref{fig:temp} but for LOS velocity, $v_{\rm LOS}$, determined from the inversion of LTE profiles ($v_{\rm LOS}^{\rm LTE}$) and NLTE profiles ($v_{\rm LOS}^{\rm NLTE}$). In the computation of relative differences $\delta v_{\rm LOS}$, pixels with $v_{\rm LOS}$ < 10\,m/s are neglected. In all the panels, positive $v_{\rm LOS}$ represents downflows.}
\label{fig:vel}
\end{figure*}

\subsection{Line-of-sight velocity}
\label{sec:vel}
The LOS velocity maps at the three nodes from the reference and test models are compared in Figure~\ref{fig:vel}. Again the two maps look similar to each other at all three nodes. \citet[][]{2013A&A...558A..20H} reported only a small difference in the statistical distribution of the Doppler shifts between the LTE and NLTE profiles. From our analysis, the histograms of the differences between the velocities in the two models, in Figure~\ref{fig:vel}, peak close to zero at all three nodes (\textit{fourth column}). The mean of the differences is between 30~ms$^{-1}$-70~ms$^{-1}$ and the standard deviation is less than 0.5~kms$^{-1}$. From the difference maps (\textit{third column}) regions of non-zero $\Delta v_{\rm LOS}$ are concentrated in the intergranular lanes, edges of granules and in the middle of magnetic elements. In the higher nodes differences of the order of 0.5~kms$^{-1}$ are seen in the middle of a few granules. The relative differences, ($\delta v_{\rm LOS}$), in the last column, are concentrated around zero in the bottom node and spread out to larger values, even up to and beyond 100\%, in the higher nodes. $\delta v_{\rm LOS}$ is large for smaller velocities. To avoid small denominator values in the computation of $\delta v_{\rm LOS}$, we have neglected pixels with $v_{\rm LOS} < 10$~ms$^{-1}$. In the top node, a large number of points is clustered around $\pm 1$~km/s, arising from intergranular lanes and granular boundaries. In these regions, inversion of NLTE profiles results in errors of up to 50\% as compared to the LTE case. When the velocities are higher, they are well constrained in the inversions and the relative differences are smaller. This is similar to the trend in the middle node.
However, the absolute error is more or less independent from the actual value. This explains why the relative error in the LTE inversions of the NLTE data increases with decreasing velocity.

 If we consider the individual intensity profiles in Figure~\ref{fig:rf_int}, at pos. 3, 4 and 5, we see a difference in the Doppler shifts, most evident at pos. 3, which is in an intergranular lane. From Table~\ref{tab:table1}, the $\Delta v_{\rm LOS}$ is large in pos. 3, 4, and 5 compared to the granular points pos. 1 and 2. At pos. 4 $\Delta v_{\rm LOS}$ also changes sign from one node to the other. The reason for this is clear from the middle panel of Figure~\ref{fig:maptau_grad} where the variation in $v_{\rm LOS}$ with $log(\tau)$ is plotted for all five points. At pos. 4, this variation is extreme with changes in direction of the flow along the vertical coordinate. Pos. 4 lies in the intergranular lane next to a magnetic element, \citep[similar to a flux sheet described in][see Figure~\ref{fig:cont1}]{2015A&A...582A.101H}, and tilting of such regions into the granule can lead to drastic variations in the flow velocity. Gradients in velocity are known to modify the NLTE line source function \citep{1967ApJ...147.1063K, 1971ApJ...165..543K,  1971ApJ...169..157C, 1974AuJPh..27..157V}, and such scenarios are common in simulated atmospheres \citep[][Figure 10]{2015A&A...582A.101H}. Lower UV opacity due to overionization of iron atoms in combination with strong velocity gradients results in the LTE and NLTE profiles sampling different heights in the atmosphere leading to significant differences in velocity between the two cases. 

\begin{figure*}[htbp]
\centering
 \includegraphics[width=0.38\textwidth]{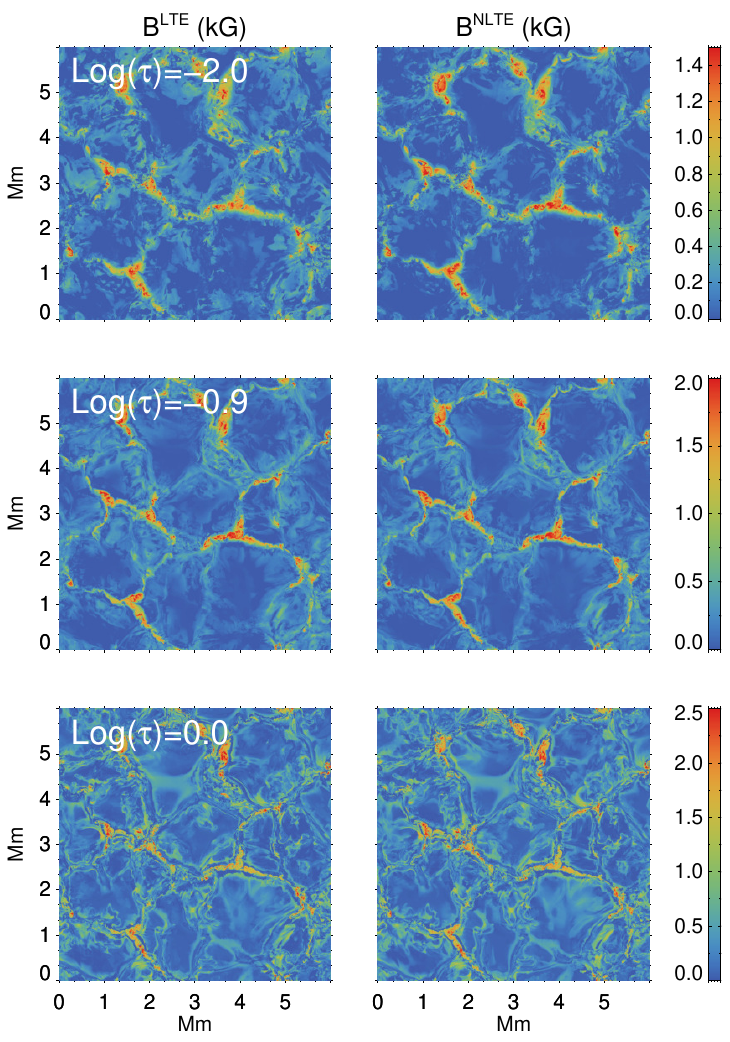}
    \includegraphics[width=0.595\textwidth]{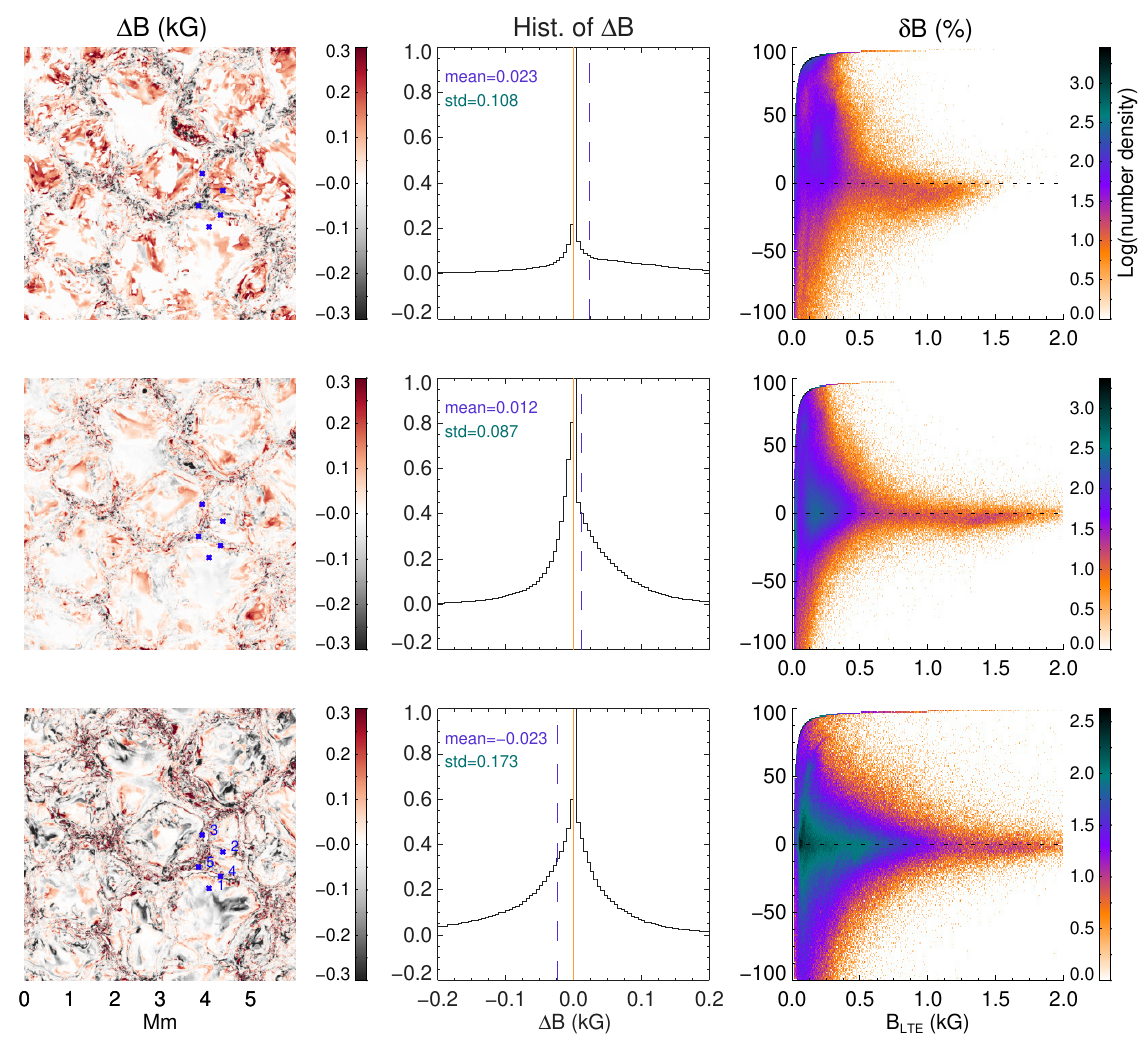}
\caption{Same as Figure~\ref{fig:temp} but for the magnetic field strength ($B$) measured by inverting the LTE profiles ($B^{\rm LTE}$) and NLTE profiles ($B^{\rm NLTE}$). In the computation of relative differences $\delta B$, pixels with $B$ < 10\,G are neglected. }
\label{fig:bfield}
\end{figure*}

\begin{figure*}[htbp]
\centering
 \includegraphics[width=0.38\textwidth]{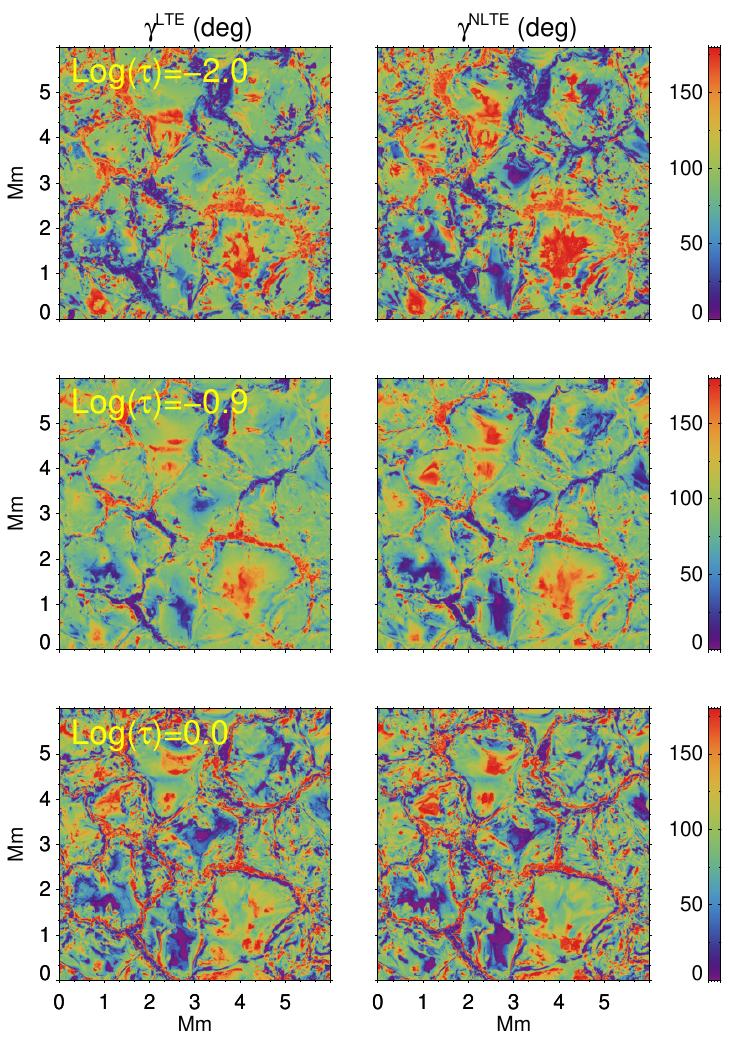}
    \includegraphics[width=0.595\textwidth]{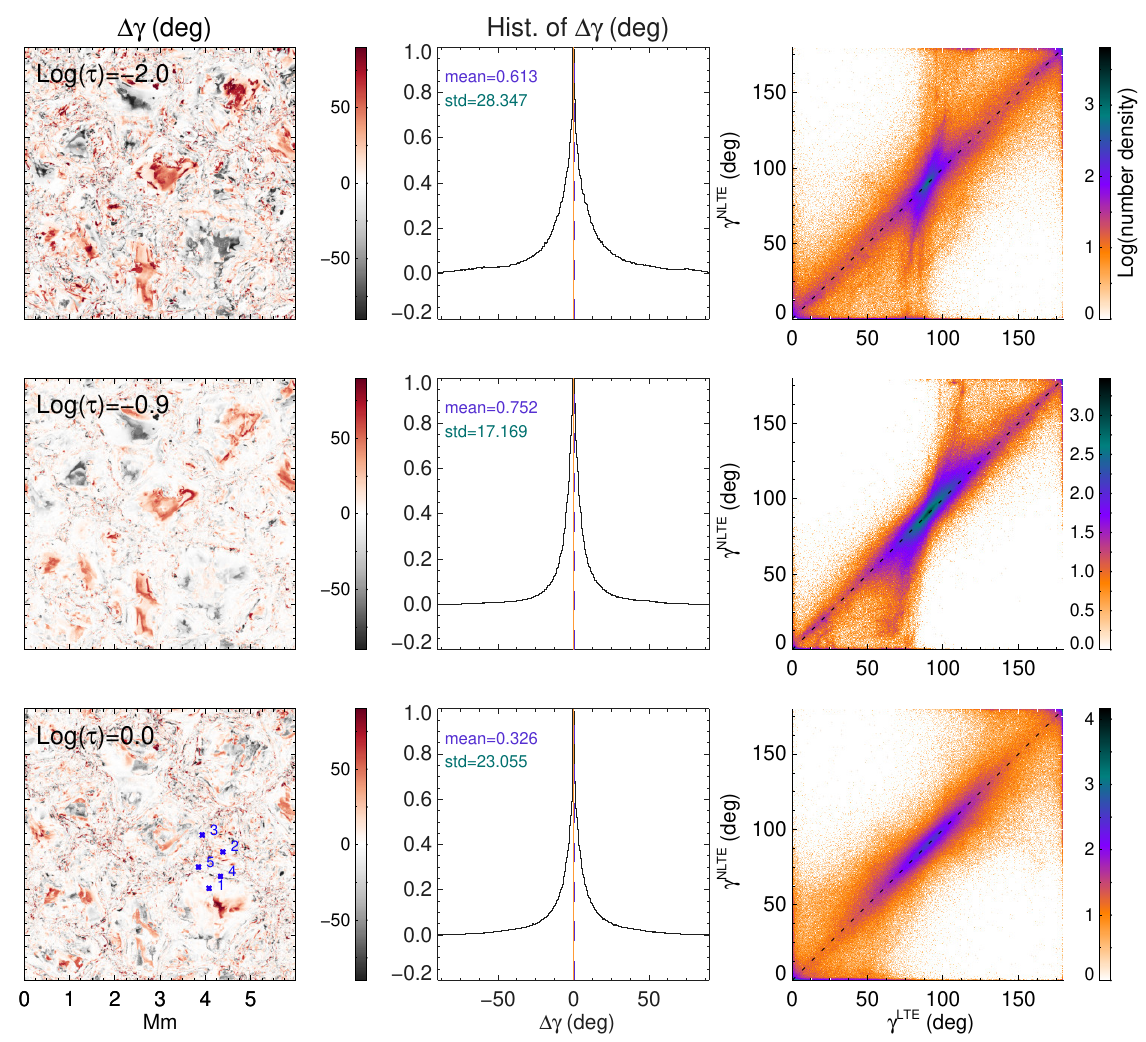}
\caption{Same as Figure~\ref{fig:temp} but for the magnetic field inclination ($\gamma$) measured from the inversion of LTE line profiles ($\gamma^{\rm LTE}$) and NLTE line profiles ($\gamma^{\rm NLTE}$).  The colorbars of the difference images in the \textit{third} column are saturated at $\pm90\degree$. The last column shows the scatter plot of $\gamma^{\rm NLTE}$ vs $\gamma^{\rm LTE}$ instead of the relative difference.}
\label{fig:gamma}
\end{figure*}

\begin{figure*}[htbp]
\centering
 \includegraphics[width=0.31\textwidth]{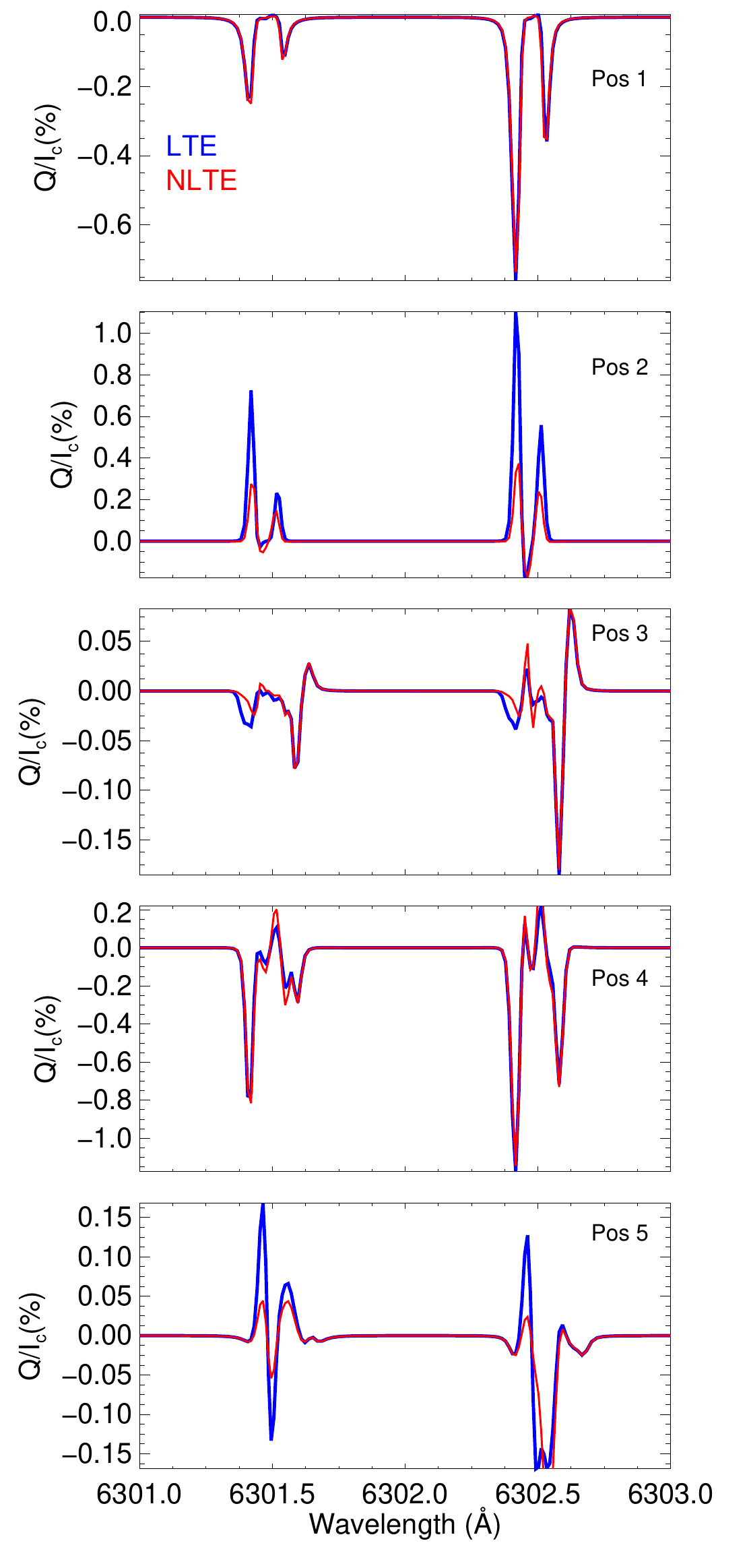}
    \includegraphics[width=0.31\textwidth]{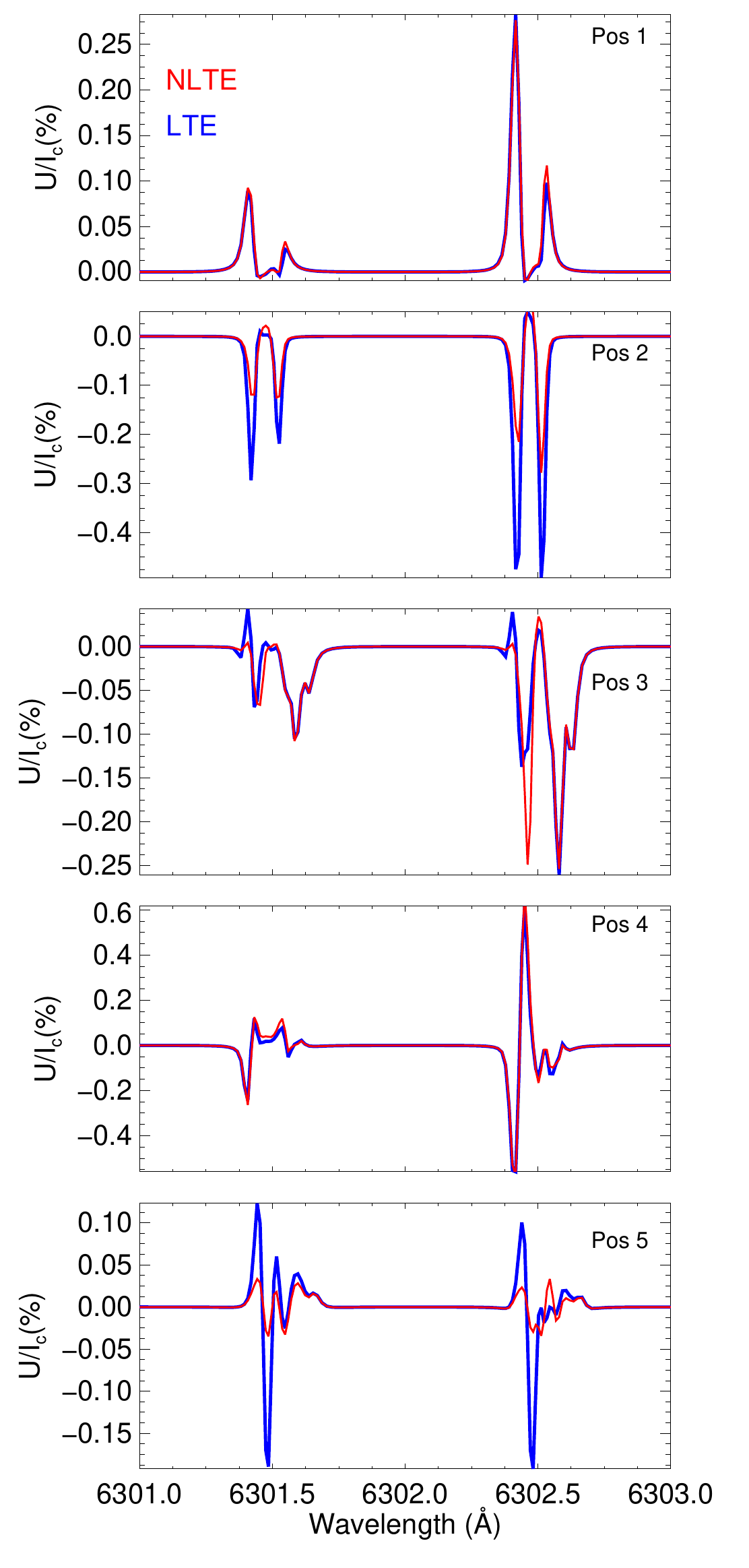}
    \includegraphics[width=0.31\textwidth]{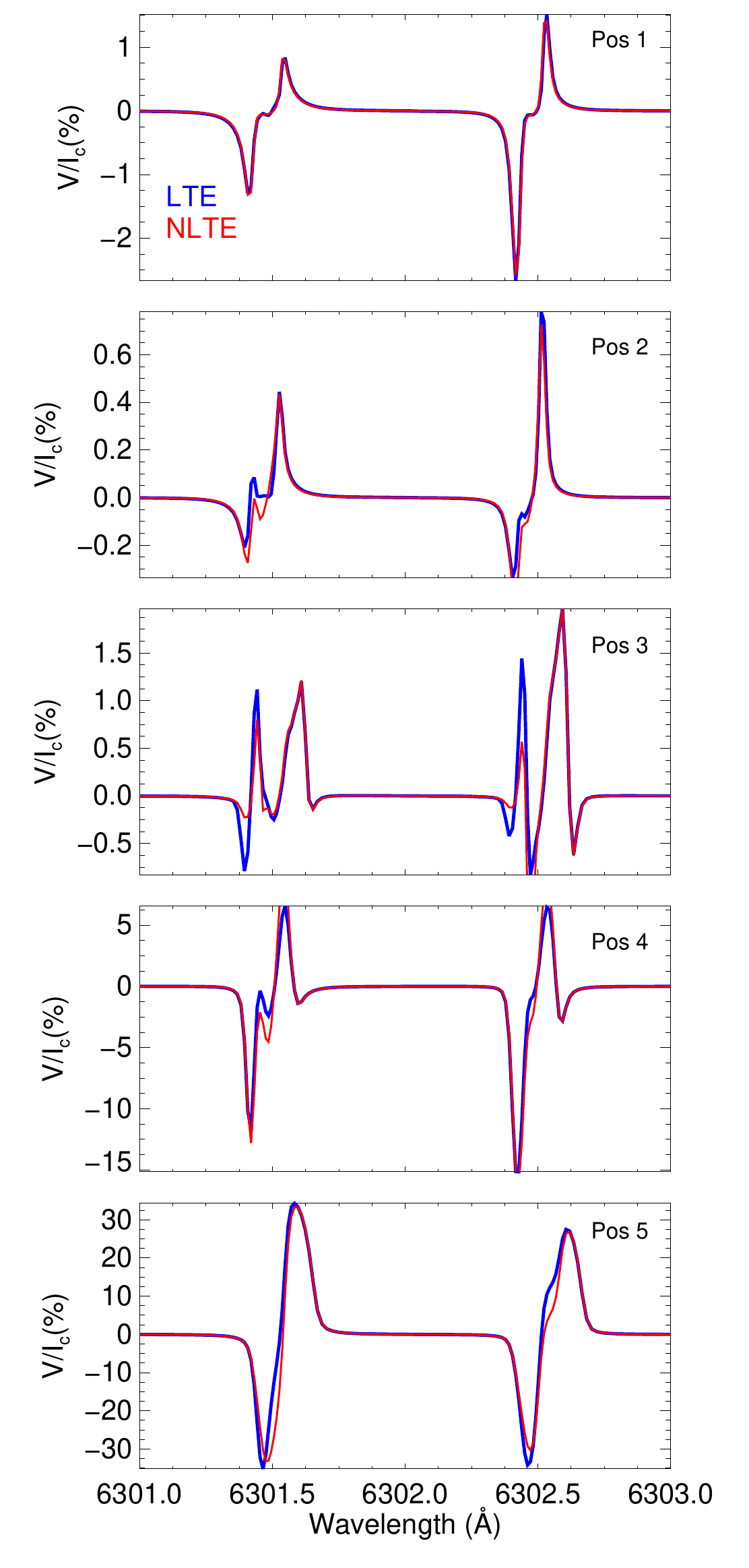}
\caption{The $Q/I_c, U/I_c$ and $V/I_c$ profiles computed in LTE (blue) and NLTE (red). They are shown at pos 1-5 marked in Figure~\ref{fig:cont1}.}
\label{fig:quv}
\end{figure*}

\subsection{Magnetic field}
\subsubsection{Strength }
\label{sec:bfield}
\citet[][]{2012A&A...547A..46H} investigated the NLTE effects in flux sheets and discussed their influence on the Stokes Zeeman profiles computed using 1D and 3D radiative transfer with a simple model and no inversions. The authors found that the field strengths recovered from the 1D-NLTE profiles of the Fe~{\sc i} 5250\,\AA{} line are higher than those from LTE. In the present paper we use the two lines at 6300\,\AA{}, which are formed slightly higher in the atmosphere than the 5250\,\AA{} line, and thus are more sensitive to NLTE effects. Among the two lines at 6300\,\AA{}, the 6301.5\,\AA{} line is formed higher but has a smaller Land\'{e} g-factor (g=1.67), whereas the 6302.5\,\AA{} line forms $\approx 100$\,km lower, but is more sensitive to the magnetic field (g=2.5). From the intensity profiles in Figure~\ref{fig:rf_int} at pos. 5, which corresponds to a point in a magnetic element, the 6302.5\,\AA{} line is completely split, with the LTE profiles being deeper and stronger ($\delta E$>0) than the NLTE profiles. we recover higher field strengths from the NLTE profiles as compared to the LTE ones, in two out of three nodes in the magnetic element (see Table~\ref{tab:table1}). 

The errors in the magnetic field strength determination due to the neglect of NLTE effects, across different structures in an MHD cube is shown in Figure~\ref{fig:bfield}. The magnetic field strength maps at all three nodes are, overall, similar in both reference and test models. The histograms of the differences (\textit{fourth column}) are highly non-Gaussian with narrow cores, broad wings with mean values less than 50~G. The standard deviation of the distribution is of the order of 100~G, with the main contribution coming from the wings of the distribution. From the relative difference, $\delta B$ plotted in the last column, the scatter is large for field strengths below 500\,G which correspond to the weak fields in and around the granules. Here, due to the fields being weak, the polarization signals are quite small, making it difficult to constrain the magnetic field. From the difference maps in \textit{third column}, $\Delta B$ is non-zero in the intergranular lanes, edges of granules and in regions with a strong magnetic field. Also from Table~\ref{tab:table1}, at pos. 3, 4 and 5, which are chosen from these three regions, not only is $\Delta B$ large, $\delta B$ is large as well (seen from the LTE values in parentheses). The MHD cube used here has large magnetic flux concentrations between granules (flux sheets lying in lanes) and at the intersections of multiple lanes (more tube-like structures). 
In these regions, neglecting the NLTE effects results in retrieval of stronger fields. Thus the points in the density plots are shifted slightly towards the negative $\delta B$, especially for fields greater than 500~G, particularly in the top node.  Figure~\ref{fig:maptau_grad} shows that, strong magnetic field gradients are present at the five chosen spatial points, which contributes to the strengthening of NLTE effects. 
The errors are smallest for the central node, which is also the one that is best constrained in the inversions. The qualitative picture is, however, reasonably consistent between nodes: fields are overestimated in lanes and in magnetic elements, but are underestimated at the edges of granules.

\subsubsection{Inclination}
Neglecting the NLTE effects introduces errors in the determination of the magnetic field inclination as well. In Figure~\ref{fig:gamma} we estimate the magnitude of this error by comparing the inclination in the test model with the reference model. Large patches of non-zero $\Delta \gamma$ are seen in the top and the central nodes in the middle of granules (\textit{third column}). In the last column, unlike in Figures~\ref{fig:temp}, \ref{fig:vel}, and \ref{fig:bfield}, we show a scatter plot of the inclination in the test model vs inclination in the reference model, instead of the relative difference. In the top and the central nodes, the scatter increases as the field becomes more horizontal. These come from regions of low (< 200\,G) magnetic field strength (see Figure~\ref{fig:bfield}). Pixels with larger $\Delta \gamma$ are seen along the intergranular lanes as well where magnetic fields are strong and vertical but the main departures are seen in the granules. Of the three nodes the standard deviation of the histogram of $\Delta \gamma$ is the smallest at the central node as it is better constrained in the inversions. The mean of this histogram is less than $1\degree$ in all three nodes.

When we examine the Stokes $Q,U$ and $V$ profiles at the five selected spatial positions in Figure~\ref{fig:quv}, we find big differences between the LTE and NLTE profiles. The magnitude of the error introduced by neglecting the NLTE effects at these five positions are given in Table~\ref{tab:table1}. The difference $\Delta \gamma$ is large at positions 1, 2, 3 and 4, especially in the top node. At pos. 1 and 2, the magnetic field from the reference model is weak and nearly horizontal. The $Q, U$ signals are small and it is difficult to constrain the magnetic field inclination. This also contributes to $\Delta \gamma$ being large. At pos. 4 and 5., the field   (Table~\ref{tab:table1}) is strong and vertical in the reference model. Here the $V$ signals are strong and $\Delta \gamma$ is also smaller compared to the other positions. The gradients in magnetic field strength, inclination, LOS velocity which result in multi-lobed asymmetric $V$ profile, enhance the NLTE effects.

\begin{figure*}[htbp]
\centering
     \includegraphics[width=0.258\textwidth]{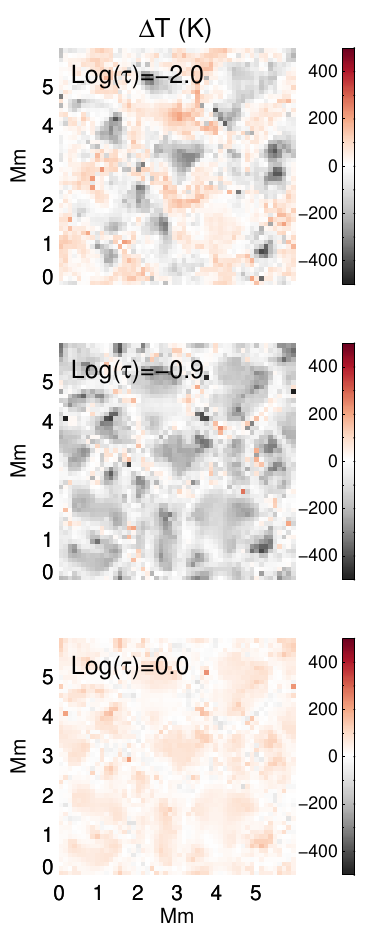}
    \includegraphics[width=0.22\textwidth]{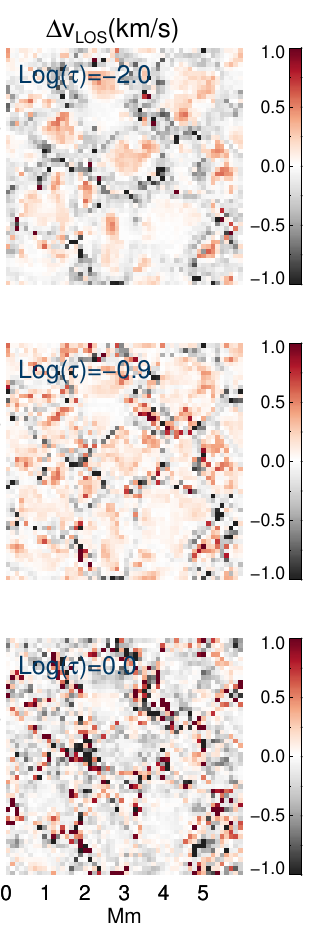}
    \includegraphics[width=0.22\textwidth]{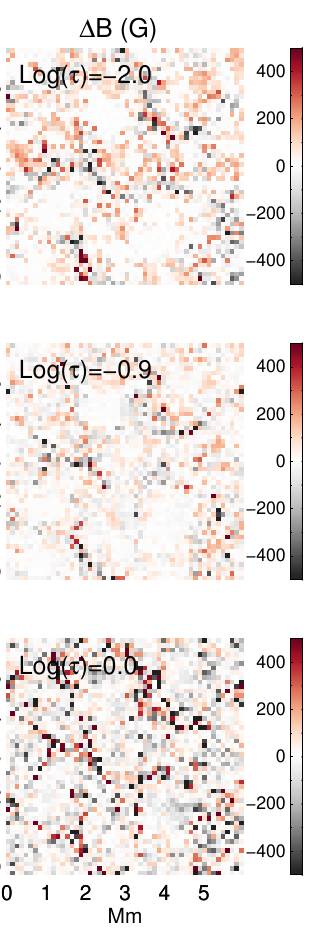}
    \includegraphics[width=0.22\textwidth]{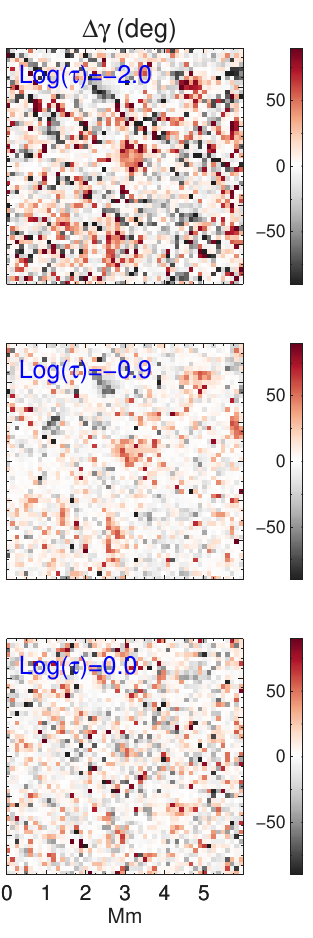}
\caption{Maps showing the differences in retrieved temperature ($\Delta T$), LOS velocity ($\Delta v_{\rm LOS}$), magnetic field strength ($\Delta B$) and inclination ($\Delta \gamma$) from the inversion of the LTE profiles and NLTE profiles. Before inverting, the profiles are rebinned to reduce the spatial resolution. The three maps for each physical parameter are for the three nodes $log(\tau)=-2.0, -0.9$ and $0.0$, used in inversion.} \label{fig:rebin}
\end{figure*}
\begin{figure*}[htbp]
    \centering
        \includegraphics[width=0.8\textwidth]{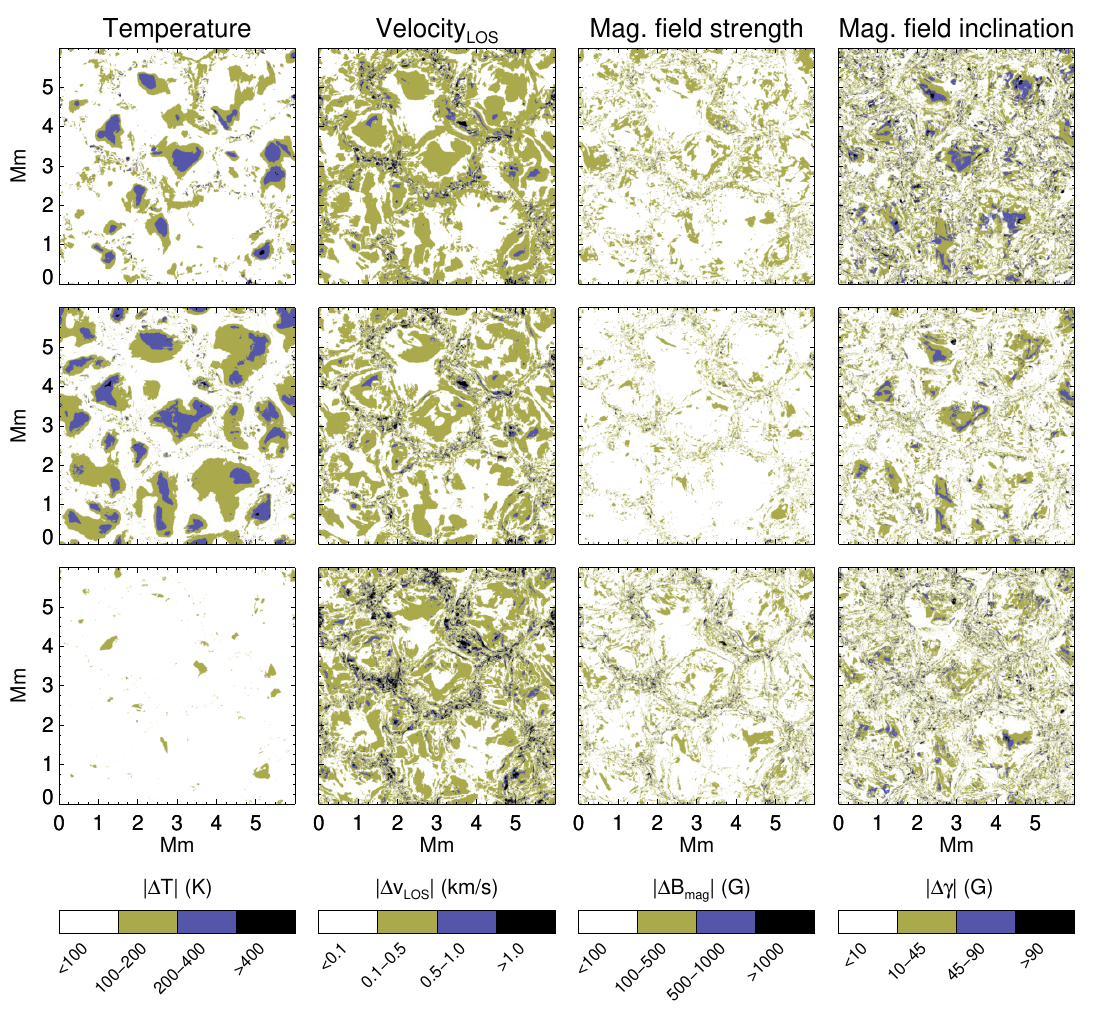}
    \caption{Maps of the regions from where the pixels in the four ranges in Table~\ref{tab:table2} originate. This is shown for $|\Delta T|, |\Delta v_{\rm LOS}|$, $|\Delta B|$ and $|\Delta \gamma|$, at all three nodes. }
    \label{fig:diff_range_maps}
\end{figure*}

\begin{table*}[htbp]
    \centering
        \caption{Percentage of pixels with absolute values of $\Delta T$, $\Delta v_{\rm LOS}$, $\Delta B$ and $\Delta \gamma$ in the specified ranges, at both original and reduced spatial scale.}
    \begin{tabular}{cccccc}
    \hline
    & $log(\tau)$ & < 100\,K & 100\,K--200\,K & 200\,K--400\,K & >400\,K\\
    \hline
        \multirow{3}{*}{$|\Delta T|$ (\%)}\\
    &-2.0 & 76.4 & 18.7 & 4.8 & 0.1  \\
    &-0.9 & 56.4 & 32.2 & 11.2 & 0.1 \\
    & 0.0 & 97.8 & 2.2 & -- & --  \\
    \hline
    \multirow{3}{*}{$|\Delta T|$ (\%)}\\
    &-2.0 & 76.8 & 18.0 & 5.0 & 0.04  \\
    (rebinned)&-0.9 & 53.6 & 34.0 & 12.0 & 0.4 \\
    & 0.0 & 96.7 & 3.1 & 0.1 & 0.04  \\
    &  & & & &\\
         \hline
         & $log(\tau)$ & < 100\,ms$^{-1}$ & 100\,ms$^{-1}$--500\,ms$^{-1}$ & 500\,ms$^{-1}$--1\,kms$^{-1}$ & >1\,kms$^{-1}$\\
    \hline
    \multirow{3}{*}{$|\Delta v_{\rm LOS}|$ (\%)}\\
    &-2.0 & 42.0 & 53.1 & 3.8 & 1.0 \\
    &-0.9 & 49.5 & 46.1 & 3.4 & 1.0 \\
    & 0.0 & 41.0 & 49.6 & 6.5 & 2.8 \\
     \hline
         \multirow{3}{*}{$|\Delta v_{\rm LOS}|$ (\%)}\\
    &-2.0 & 43.0 & 50.7 & 4.7 & 1.6 \\
    (rebinned) &-0.9 & 46.1 & 46.0 & 6.1 & 1.8 \\
    & 0.0 & 37.4 & 48.4 & 8.6 & 5.5 \\
    & & & & & \\
    \hline
    & $log(\tau)$ & < 100\,G & 100\,G--500\,G & 500\,G--1\,kG & >1\,kG\\
    \hline
    \multirow{3}{*}{$|\Delta B|$ (\%)}\\
    &-2.0 & 76.2 & 23.2 & 0.4 & 0.05  \\
    &-0.9 & 87.5 & 12.1 & 0.3 & 0.07  \\
    & 0.0 & 74.1 & 23.6 & 1.8 & 0.5  \\
         \hline
     \multirow{3}{*}{$|\Delta B|$ (\%)}\\
    &-2.0 & 72.6 & 26.5 & 0.9 & 0.04  \\
    (rebinned) &-0.9 & 81.5 & 17.8 & 0.6 & 0.11  \\
    & 0.0 & 68.0 & 28.2 & 3.3 & 0.6  \\
         \hline
             & $log(\tau)$ & $< 10\degree$ & $10\degree-45\degree$ & $45\degree-90\degree$ & $>90\degree$\\
    \hline
    \multirow{3}{*}{$|\Delta \gamma|$ (\%)}\\
    &-2.0 & 50.6 & 37.9 & 10.3 & 1.2  \\
    &-0.9 & 69.2 & 26.8 & 3.9 & 0.1  \\
    & 0.0 & 60.3 & 33.0 & 5.9 & 0.8  \\
         \hline
     \multirow{3}{*}{$|\Delta \gamma|$ (\%)}\\
    &-2.0 & 38.7 & 39.8 & 17.7 & 3.8  \\
    (rebinned) &-0.9 & 59.5 & 34.7 & 5.3 & 0.5  \\
    & 0.0 & 52.2 & 38.6 & 8.1 & 1.1  \\
         \hline
    \end{tabular}
\label{tab:table2}
\end{table*}

\section{Inversion of rebinned profiles}
\label{sec:rebin}
The high resolution 3D MHD cube used for the analyses in the present paper was generated on a very fine grid with a spacing of 5.8km/pixel. This is much smaller than the spatial scales that can be resolved with current instrumentation. The \textit{Hinode} satellite, which is extensively used for solar photospheric observations of the 6300\,\AA{} lines, has a resolution of $0.16^{\prime\prime}$/pixel, which corresponds to about 116 km/pixel, 20 times larger than the resolution of our MHD cube. Many of the locations where the differences between the reference and test models were discussed in the previous sections lie within small-scale fine structures, resolved in the MHD cube, but not in any observations. It is therefore important to know if the errors in the test model discussed in Section~\ref{sec:atm_comp} persist when the resolution is reduced, for example, to roughly match that of \textit{Hinode}.

To test this, ideally, the Stokes profiles should be convolved with the point spread function (PSF) of the instrument, the profiles rebinned and noise added. Since our aim is to test if spatial averaging weakens the NLTE effects as discussed in  \citet[][]{2001ApJ...550..970S}, we only rebin the profiles to match the spatial scales observed by \textit{Hinode} and do not convolve them with a PSF or add noise. 
The Stokes profiles computed in LTE and NLTE, respectively, are then inverted in LTE using the exact same setting as before. The temperature, LOS velocity, magnetic field strength, and inclination maps from the inversion of LTE and NLTE profiles are compared in a way similar to that used for the full resolution case, discussed in the previous sections. In Figure~\ref{fig:rebin}, we show only the difference maps for the four atmospheric parameters. Comparing these maps with the original resolution maps in Figures~\ref{fig:temp}, \ref{fig:vel}, \ref{fig:bfield}, and \ref{fig:gamma} shows a striking similarity. For the temperature, as in the original resolution case, differences are seen mostly in the granules with $T_{\rm LTE} < T_{\rm NLTE}$ in many of them, this of course also depends on the node. A few exceptions are seen at $log(\tau)=-2.0$ in locations with weaker temperature gradients, as discussed in detail in Section~\ref{sec:temp_nodes12}. 

For the LOS velocity and magnetic field strength, the locations with significant differences are more pronounced and appear to be clustered in the intergranular lanes and regions with strong magnetic fields. These are the regions with strong vertical gradients and the Stokes profiles, even after rebinning, contain these imprints, which are reflected in the corresponding inverted atmospheres. Differences in velocity are seen also in the middle of granules, like in Figure~\ref{fig:vel}. The magnitude of the errors is as large as in the original case. 

In the maps showing errors in the determined magnetic field inclination, we see a large number of pixels with increased errors or differences, especially in the top node. The errors are smallest in the middle node just as in the full resolution case.

\section{Statistics of the differences}
To further quantify the NLTE effects, we divide the absolute difference $|\Delta T|, |\Delta v_{\rm LOS}|$, $|\Delta B|,$ and $|\Delta \gamma|$ into four different ranges. In Table~\ref{tab:table2}, we indicate the percentage of pixels in each range and in Figure~\ref{fig:diff_range_maps}, we map the four ranges to the regions where they originate. In temperature maps, the majority of the pixels have $|\Delta T|$ < 200\,K in the top two nodes. At the same time, pixels with $|\Delta T|$ between 200\,K-400\,K are non-negligible and are concentrated in the middle of granules. Nearly 5\% and 11\% of the pixels fall in this range in the top and middle node, respectively. NLTE effects from both line core and the wings contribute to the temperature determination in the middle node (see response functions in Figure~\ref{fig:rf_int}) resulting in a higher percentage of pixels with large $|\Delta T|$. In the bottom node, $|\Delta T|$ < 100\,K with a few pixels here and there having larger differences. Ideally $|\Delta T|$ should be zero but, due to a poor separation of the continuum from the line wings, it is non-zero, as already discussed in Section~\ref{sec:temp_node0}. 

Regarding the determined LOS velocity, the majority of the pixels have differences $ < 0.5$\,kms$^{-1}$. Less than $10\%$ have $|\Delta v_{\rm LOS}|$ between 0.5 kms$^{-1}$ and 1.0 kms$^{-1}$. A small fraction $\approx 1\%$ of the pixels show differences larger than 1\,kms$^{-1}$. As discussed in Section~\ref{sec:vel} and as can be seen from Figure~\ref{fig:diff_range_maps}, pixels with large differences are found along the intergranular lanes. 
Nearly all pixels, at all three nodes, have differences less than 500\,G in the determined magnetic field strength. Only in the bottom node, $\approx 2\%$ of the pixels have $|\Delta B|>$1\,kG. These pixels are found in small groups, sparsely distributed across the whole region (see Figure~\ref{fig:diff_range_maps}). 

Although the error in the magnetic field inclination is less than $10\degree$ in most of the pixels, a significant fraction of pixels ($30\%-40\%$) have errors between $10\degree-45\degree$, in all three nodes. In the top node, nearly $10\%$ of the pixels have errors greater than $45\degree$. These problematic pixels are found in the centers of the granules where the magnetic field strength is low. When the spatial resolution is reduced, the percentage of pixels with error $>45\degree$ increases to $~22\%$. 

An important result from Table~\ref{tab:table2} is that the percentage of pixels with quantitative errors in $T, v_{\rm LOS},B$ and $\gamma$ remains the same, or even increases for $\gamma$, after spatially averaging the Stokes profiles. Although earlier papers such as \citet[][]{2001ApJ...550..970S} predicted the spatial averaging to weaken the NLTE effects, our investigations show that they do survive. From  Figure~\ref{fig:diff_range_maps}, the errors in test model are not just found in a few isolated pixels but are seen as patches. Spatial averaging of the Stokes profiles does not cancel out these errors, at least when the horizontal transfer effects are not taken into account. 

\section{Summary and conclusions}

The influence of NLTE conditions on the formation of photospheric iron lines has been well-known for several decades. However, due to the complexities involved in NLTE radiative transfer theory, these lines are often assumed to be formed in LTE, in particular when doing inversions of Stokes profiles. When  inverting their Stokes profiles, not only are the effects of NLTE neglected, but also those of 3D radiative transfer for the same reasons. In continuation to the work of \citet[][]{2012A&A...547A..46H, 2013A&A...558A..20H, 2015A&A...582A.101H},  we investigate how the assumption of LTE in the inversions of NLTE iron profiles introduces errors in the inferred atmospheres. In this first step, we assume that the Fe~{\sc i} 6301.5\,\AA{} and 6302.5\,\AA{} lines are formed in 1D NLTE. We compare the atmosphere determined from the inversion of line profiles computed in NLTE with the atmosphere from the inversion of lines formed in LTE. The latter is used as a reference model. The atmospheric model chosen for the inversion is a simple 3-node atmosphere which can be inadequate to reproduce extreme variations of physical quantities in the MHD cube. Such artifacts of the inversion can also contribute to the scatter in relative differences and lead to uncertainties. However, the strategy of using an inverted atmosphere as a reference model instead of the actual MHD cube is expected to mitigate this issue.

While most of the previous studies have focused on discussing how the NLTE conditions can alter the measured temperature, we  examine the LOS velocity and magnetic field as well.
In the inferred temperature, we find that neglecting NLTE can result in departures of up to 13\% in both top and central inversion nodes. In the LOS velocity and magnetic field strength, we find relative difference as high as 50\% or more in multiple nodes. A large fraction of the pixels have errors up to $45\degree$ in the determined magnetic field inclination. It is not possible to generalize whether neglecting the NLTE effects results in an overestimation or an underestimation of a particular parameter, since this depends on the vertical gradients in the region of interest and the amount of UV radiation from the layers below. 

We chose five representative spatial positions to compare the LTE and NLTE Stokes profiles. We tried to correlate the errors in the inferred atmosphere at these locations with the differences observed in the Stokes profiles. 
In the intergranular lane and in the magnetic elements, the LOS velocity is off by$~500$\,m/s  compared to the reference model. This can be a significant error when the LOS velocity itself is of the order of a few 100\,ms$^{-1}$ such as in the quiet Sun bright points \citep{Riethm_ller_2010,Romano2012}. Errors in magnetic field strength and inclination are evident in both weak and strong magnetic field regions. Given that the 6300\,\AA{} are often used in the construction of 1D model atmospheres based on LTE inversions \citep[e.g.,][]{Bellot_Rubio_1997, 2002A&A...391..331B,2002A&A...385.1056B,2013A&A...554A.116F, 2017ApJ...841..115C}, neglecting NLTE effects will introduce errors in these models.

A large scatter in the relative difference ($\delta B$) is observed at all three nodes in the test model in Figure~\ref{fig:bfield}. The small-scale weak magnetic field close to the photosphere is difficult to constrain with the simple inversion model as the polarization signals are small. While this contributes to the observed scatter in the test model, the role of NLTE effects cannot be ruled out. This is accompanied by errors of magnitude $10\degree-45\degree$ (Figure~\ref{fig:gamma}) in the determined magnetic field inclination. LTE inversions of observations at Fe~{\sc i} 6300\,\AA{} lines are often used for measuring the strength, inclination of internetwork magnetic field and their variation with height \citep[e.g.,][]{2002ApJ...573..431L, 2007ApJ...670L..61O, 2012A&A...547A..89B, 2016A&A...593A..93D}. Understanding the errors due to the neglect of NLTE effects is important for the interpretation of such measurements.

While most the papers discussing NLTE effects in iron lines mainly focus on the quiet Sun, a detailed study in active regions is missing. The only known work is by \citet[][]{1997ASPC..118..207S}, where the authors conclude that neglecting the NLTE effects can lead to an under-estimation of sunspot temperature. Observations at the Fe~{\sc i} 6300\,\AA{} line pair are extensively used in understanding the physical properties of active regions such as sunspots \citep[e.g.,][]{2008ApJ...678L.157R,2014A&A...568A..60L, 2015A&A...583A.119T,2017A&A...599A..35J,2018A&A...611L...4J} and plages \citep[e.g.,][]{2015A&A...576A..27B} under the assumption that the lines are formed in LTE. How important are NLTE effects in active regions? How do they affect the temperature, velocity and magnetic field? These questions can be answered by carrying out an investigation similar to the one presented in the current paper using a sunspot simulation from \citet[][]{2012ApJ...750...62R}. This will be the subject of an interesting follow-up work.

With the availability of a few NLTE inversion codes such as NICOLE, SNAPI \citep{2018A&A...617A..24M}, and STiC \citep{2018arXiv181008441D}, it should now become possible to carry out the complex NLTE inversion of the photospheric iron lines at least using 1D radiative transfer. Finally, according to  \citet[][]{2015A&A...582A.101H}, the effects of horizontal radiative transfer can intensify or weaken the effects of NLTE. How the combination of horizontal radiative transfer with NLTE affects the inverted atmosphere will be investigated in a future publication. 

\begin{acknowledgements}
We thank L. P. Chitta and I. Mili\'{c} for comments on the manuscript. This project has received funding from the European Research Council (ERC) under the European Union's Horizon 2020 research and innovation programme (grant agreement No. 695075). This work has also been partially supported by the BK21 plus program through the National Research Foundation (NRF) funded by the Ministry of Education of Korea. This research has made use of NASA’s Astrophysics Data System.
\end{acknowledgements}


\end{document}